%% file: main.tex
\documentclass[runningheads]{llncs}

\makeatletter
\RequirePackage[bookmarks,unicode,colorlinks=true,breaklinks=true]{hyperref}%
   \def\@citecolor{blue}%
   \def\@urlcolor{blue}%
   \def\@linkcolor{blue}%

\def\orcidID#1{\smash{\href{http://orcid.org/#1}{\protect\raisebox{-1.25pt}{\protect\includegraphics{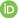}}}}}
\makeatother

\usepackage{graphicx} 
\usepackage{amsmath}
\allowdisplaybreaks
\usepackage{amsfonts}
\usepackage{amssymb}
\usepackage{tikz}
\usetikzlibrary{arrows.meta,automata}
\usepackage{subcaption}
\usepackage{carshapes}
\usepackage{pgfplots}
\usepackage{forest}

\usepackage{algorithm}
\usepackage{algcompatible}
\usepackage{nccmath}
\usepackage{multirow}
\usetikzlibrary{decorations,decorations.markings,chains,shadows,shapes.geometric,shapes.multipart} 
\usetikzlibrary{positioning,arrows,backgrounds}
\usetikzlibrary{patterns}
\usepackage{listings}
\usepackage{scalerel}

\pgfkeys{/tikz/.cd,
    street mark distance/.store in=\StreetMarkDistance,
    street mark distance=10pt,
    street mark step/.store in=\StreetMarkStep,
    street mark step=1pt,
    /pgf/number format/.cd,fixed relative,precision=2
}

\pgfdeclaredecoration{street mark}{initial}
{%
\state{initial}[width=\StreetMarkStep,next state=cont] {
    \pgfmoveto{\pgfpoint{\StreetMarkStep}{\StreetMarkDistance}}
    \pgfpathlineto{\pgfpoint{0.3\pgflinewidth}{\StreetMarkDistance}}
    \pgfcoordinate{lastup}{\pgfpoint{1pt}{\StreetMarkDistance}}
    
  }
  \state{cont}[width=\StreetMarkStep]{
     \pgfmoveto{\pgfpointanchor{lastup}{center}}
     \pgfpathlineto{\pgfpoint{\StreetMarkStep}{\StreetMarkDistance}}
     \pgfcoordinate{lastup}{\pgfpoint{\StreetMarkStep}{\StreetMarkDistance}}
  }
  \state{final}[width=\StreetMarkStep]
  {
    \pgfmoveto{\pgfpointdecoratedpathlast}
  }
}

\newcommand\doverline[1]{\ThisStyle{%
  \setbox0=\hbox{$\SavedStyle\overline{#1}$}%
  \ht0=\dimexpr\ht0-.15ex\relax
  \overline{\copy0}%
}}

\newcommand{\custompath}{(0,0) to[out=90,in=270] (0,4)}

\lstdefinelanguage{Prism} {
    keywords=[0]{const,global,bool,int,double,dtmc,ctmc,mdp,csg,rewards,endrewards,formula,module,endmodule,player,endplayer,init,true,false},    basicstyle=\linespread{1.0}\scriptsize\ttfamily\color{black},
    keywordstyle=[0]\ttfamily\color{purple},
    keywordstyle=[1]\ttfamily\color{teal},
    comment=[l][\color{gray}]{//},
    sensitive=true,
    frame=single,	
    numbers=left,
    numberstyle=\tiny\color{gray},
    morestring=[b]"
}

\input{macros-dave.tex}

\begin{document}

\title{Expectation vs. Reality: \\ Towards Verification of Psychological Games}
\titlerunning{Expectation vs. Reality: Towards Verification of Psychological Games}
\author{Marta Kwiatkowska\inst{1}\orcidID{0000-0001-9022-7599}
\and Gethin Norman\inst{1,2}\orcidID{0000-0001-9326-4344}
\and \\ David~Parker\inst{1}\orcidID{0000-0003-4137-8862} 
\and Gabriel Santos\inst{1} \orcidID{0000-0002-6570-9737}
}
\institute{Department of Computer Science, University of Oxford, Oxford, UK \email{\{marta.kwiatkowska,david.parker,gabriel.santos\}@cs.ox.ac.uk}
\and
School of Computing Science, University of Glasgow, Glasgow, UK
\email{gethin.norman@glasgow.ac.uk}}

\maketitle

\begin{abstract}
Game theory provides an effective way to model strategic interactions among rational agents. In the context of formal verification, these ideas can be used to produce guarantees on the correctness of multi-agent systems, with a diverse range of applications from computer security to autonomous driving.
Psychological games (PGs) were developed as a way to model and analyse agents with belief-dependent motivations, opening up the possibility to model how human emotions can influence behaviour. In PGs, players' utilities depend not only on what \emph{actually} happens (which strategies players choose to adopt), but also on what the players had \emph{expected} to happen (their belief as to the strategies that would be played).
Despite receiving much attention in fields such as economics and psychology, very little consideration has been given to their applicability to problems in computer science, nor to practical algorithms and tool support. In this paper, we start to bridge that gap, proposing methods to solve PGs and implementing them within PRISM-games, a formal verification tool for stochastic games. We discuss how to model these games, highlight specific challenges for their analysis and illustrate the usefulness of our approach on several case studies, including human behaviour in traffic scenarios.
\end{abstract}

\input{introduction}

\input{preliminaries}
\input{modelchecking}
\input{csgs}

\input{experiments}
\input{conclusions}

\bibliographystyle{splncs04.bst}
\bibliography{bib}

\input{appendix}

\end{document}

%% file: macros-dave.tex
\renewcommand{\emptyset}{\varnothing}

\usepackage{scalerel}

\newcommand\scale[2]{\vstretch{#1}{\hstretch{#1}{#2}}}

\newcommand{\sectref}[1]{Section~\ref{#1}}

\newcommand{\figref}[1]{Figure~\ref{#1}}

\newcommand{\eqnref}[1]{(\ref{#1})}

\newcommand{\lemref}[1]{Lemma~\ref{#1}}
\newcommand{\defref}[1]{Definition~\ref{#1}}

\spnewtheorem{assumption}{Assumption}{\bfseries}{\itshape}
\newcounter{exampcount}
\newenvironment{examp}
{\refstepcounter{exampcount}
\vskip6pt\noindent
{\bf Example \arabic{exampcount}.}}
{\hfill$\blacksquare$\vskip6pt}

\newcommand{\startpara}[1]{{%
\vskip6pt\noindent
{\bf #1.}}}

\def\ra{{\rightarrow}}

\def\Nset{\mathbb{N}}
\def\Qset{\mathbb{Q}}
\def\Rset{\mathbb{R}}

\def\Eset{\mathbb{E}}

\def\ra{\rightarrow} 
\def\rmdef{\,\stackrel{\mbox{\rm {\tiny def}}}{=}}

\renewcommand{\leq}{\leqslant}
\renewcommand{\geq}{\geqslant}

\newcommand\game{{\sf C}}
\newcommand\nfgame{{\sf N}}
\newcommand\nfpgame{{\sf N_P}}
\newcommand\mgame{{\sf Z}}

\newcommand\sinit{{\bar{s}}}

\newcommand\dist{{\mathit{Dist}}}
\newcommand\Dist{{\dist}}

\newcommand\Prob{{\mathit{Prob}}}

\newcommand\strat{{\sigma}}

\newcommand{\last}{\mathit{last}}
\newcommand{\ipaths}{\mathit{IPaths}}
\newcommand{\fpaths}{\mathit{FPaths}}

\newcommand{\Belief}{{B}}
\newcommand{\probact}[1]{{p_\mathit{#1}}}

%% file: introduction.tex
\section{Introduction}

Probabilistic model checking is a well established technique for formally verifying 
computerised systems that operate in uncertain or stochastic environments. In order to verify systems comprising multiple autonomous agents
and/or those involving human interactions, various models and concepts from game theory have been adapted for probabilistic model checking.
Stochastic games, in particular, have shown to be a versatile and useful formalism to model and study situations involving collaboration or competition among agents,
successfully applied to, for example 
human-in-the-loop autonomous systems~\cite{FWHT16},
robot navigation in the presence of humans~\cite{JJK+18}
and attack-defence scenarios~\cite{ANP16}.

While traditional game theory is often used to model human decision making, it is unable to model situations such as emotional response and social norms,
where the utilities that players aim to maximise can depend on their \emph{beliefs}.
This inadequacy has been pointed out in \cite{GPS89}, which proposed the seminal model of \emph{psychological games} (PGs).
In these games, a player's utility depends not only on what \emph{actually} happens in the game (i.e., which strategies are chosen by the players), but also on what the players had \emph{expected} to happen (i.e., their belief as to the future behaviour of the other players).

This class of models makes it possible to consider different aspects that contribute to human decision-making, such as regret, trust, fear, reciprocity and fairness, and how these may influence player behaviour.
Crucially, psychological game predictions have been reproduced in human experiments, thus supporting the notion of belief-dependent motivations \cite{AN08}.
This has been of particular relevance in economics, when trying to predict and understand how people behave regarding non-material payoffs.
Naturally, though, as autonomous computerised systems become more commonplace,
ensuring that they interact safely and efficiently
with humans will also require this kind of reasoning.

In this paper, we make the first steps towards a more practical approach
to modelling and analysing PGs, and in applying them to other scenarios.
We begin with one-shot (normal form) games,
considering the \emph{normal form psychological games} (NFPGs) proposed in~\cite{GPS89}.
We work with the commonly employed solution concept of \emph{Nash equilibria} (NE),
which establishes rational strategies for a game to be those where
no player has an incentive to unilaterally deviate from their strategy.
Using the psychological extension of NE from~\cite{GPS89},
we propose an approach to finding optimal equilibria for NFPGs
using support enumeration and non-linear programming,
and highlight 
why computing equilibria for such games
is more computationally challenging.

We next investigate extensive (multi-stage) games with psychological payoffs,
under the assumption that beliefs are \emph{local} and \emph{state}-based.
We do so by considering an extension of concurrent stochastic games (CSGs)
whose reward functions can depend on both the actions taken and beliefs about those actions,
proposing a method to find equilibria for finite-horizon cumulative rewards using backward induction.
We develop prototype tool support for psychological games,
building on the PRISM-games model checker for stochastic games~\cite{KNPS20}.
Using this, we model and analyse a variety of psychological games,
notably studying human behaviour in several different traffic scenarios,
and showcase the analysis and insights made possible by our approach.

\startpara{Related work}
Psychological games were proposed in~\cite{GPS89} and shown to admit standard game-theoretic techniques such as backward induction under some restrictions. However, they assume a fixed payoff structure and do not support belief inference or updating.
\emph{Dynamic psychological games} \cite{BD09,BCD19,BD22} address some of these limitations by allowing belief update. More specifically, they remove restrictions enforced in \cite{GPS89} that make beliefs endogenous to the games, and propose a \emph{forward induction} algorithm with belief updates which allows for more sophisticated analysis. 
In \cite{Rab93}, \emph{fairness equilibria} are introduced as an extension of the framework established in \cite{GPS89}, where the payoff of each player is defined as a combination of a material payoff and a psychological payoff whose value depends on how fairly they think they are being treated. 

From an application perspective, conventional game theory has been employed to model a range of road user behaviours \cite{Elv14,KSF+23,SKC23,Tor15,MB18,AAD+24}, including merging into traffic and speed selection, and has been able to explain how informal norms of behaviour can develop among road users and be sustained even if these informal norms violate the formal regulations of the traffic code. In \cite{KSF+23}, the authors point out that autonomous agents should have inferable behaviour and model a pedestrian crossing interaction as a repeated bimatrix Stackelberg game in order to measure and establish a bound to inferability loss. A more complex interaction involving vehicles, pedestrians and cyclists at a crossing was investigated as a non-zero sum game in \cite{Tor15}, which showed that real life behaviour corresponded to an equilibrium strategy that went against Norwegian traffic laws. This example served as inspiration for \cite{MB18}, which used Bayesian games and examined possible differences in the strategies pedestrians and cyclists would be likely to adopt when considering autonomous and human drivers. 
Another example is \cite{AAD+24}, which also considers a game-theoretic stochastic model to analyse interactions among pedestrians, autonomous and regular vehicles and investigates strategies for conflict resolution in uncontrolled traffic environments based on Stackelberg equilibria.
However, to the best of our knowledge, psychological games have not been explored in road user scenarios.

The verification community has developed various software tools with support for Nash equilibria, 
such as PRALINE~\cite{BRE13}, EAGLE~\cite{TGW15} and EVE~\cite{GNP+18},
but we are aware of no tool support for psychological equilibria computation,
in either normal or extensive forms.
For probabilistic systems, model checkers such as PRISM~\cite{KNP11} and Storm~\cite{HJK+22} support a wide range of probabilistic models, with partially observable Markov decision processes providing an alternative way to reason about belief,
over (unobservable) states rather than strategies.
PRISM-games~\cite{KNPS20} provides verification and equilibria synthesis for various types of stochastic games including CSGs, but until now not for psychological variants.

%% file: preliminaries.tex
\section{Preliminaries}\label{sect:prelims}

We first recall normal form games (NFGs), over which we define Nash equilibria (NE), and then proceed by defining the psychological equivalents: normal form psychological games (NFPGs) and psychological Nash equilibria (PE). 

\startpara{Classical games}  We will write $\dist(X)$ for the set of probability distributions over a finite set $X$.

\begin{definition}[Normal form game] A (finite, $n$-person) \emph{normal form game} (NFG) is a tuple $\nfgame = (N,A,u)$ where:
\begin{itemize}
\item $N=\{1,\dots,n\}$ is a finite set of \emph{players};
\item $A = A_1 \times \cdots \times A_n$, $A_i$ is a finite set of \emph{actions} available to player $i \in N$ and $A_i \cap A_j = \emptyset$ for $i \neq j \in N$;
\item $u = (u_1,\dots,u_n)$ and $u_i \colon A \rightarrow \Qset$ is a \emph{utility function} for player $i \in N$.
\end{itemize}
\end{definition}
In an NFG, each player $i \in N$ simultaneously chooses an action $a_i \in A_i$ from their action set
and each player $j$ then receives utility $u_j(a_1,\dots,a_n)$.
Two-player NFGs are often called \emph{bimatrix games} since they can be represented by two matrices $\mgame_1, \mgame_2 \in \Qset^{l \times m}$ where $A_1 = \{a_1,\dots,a_l\}$, $A_2 = \{b_1,\dots,b_m\}$, $\mgame_1(i,j) = u_1(a_i,b_j)$ and $\mgame_2(i,j) = u_2(a_i,b_j)$. 
Below, we assume a fixed NFG $\nfgame = (N,A,u)$.

\begin{definition}[Strategy and strategy profile] 
A \emph{(mixed) strategy} for player $i$ of NFG $\nfgame$ is a distribution $\sigma_i \in \dist(A_i)$,
specifying the probability of choosing each action in its action set.
A \emph{strategy profile} of $\nfgame$ (or just \emph{profile})
is a tuple $\sigma = (\sigma_1,\dots,\sigma_n)$ of strategies for all players.
\end{definition}
The \emph{expected utility} of player $i$ under strategy profile $\sigma = (\sigma_1,\dots,\sigma_n)$ is: 
\[ \begin{array}{c}
u_i(\sigma) \ \rmdef \ \sum_{(a_1,\dots,a_n) \in A} u_i(a_1,\dots,a_n) \cdot \left( \prod_{j=1}^n \sigma_j(a_j) \right) \, .
\end{array} \]
We let $\Sigma^i_\nfgame = \dist(A_i)$ denote the set of all player $i$ strategies,
$\smash{\Sigma_\nfgame = \prod_{i \in N} \Sigma^i_\nfgame}$ the set of all strategy profiles
and $\smash{\Sigma^{-i}_\nfgame = \prod_{j \neq i} \Sigma^j_\nfgame}$ the set of strategy tuples for all players except $i$.
For strategy profile $\sigma = (\sigma_1,\dots,\sigma_n)$ and player $i$ strategy $\sigma_i'$, we define the strategy tuple $\sigma_{-i} = (\sigma_1,\dots,\sigma_{i-1},\sigma_{i+1},\dots,\sigma_n)$ and strategy profile $\sigma_{-i}[\sigma_i'] = (\sigma_1,\dots,\sigma_{i-1},\sigma_i',\sigma_{i+1},\dots,\sigma_n)$.

\begin{definition}[Support]\label{def:support}
The \emph{support} $Q_i \subseteq A_i$ of a strategy $\sigma_i$ for player $i$ is the set of actions it chooses with positive probability, i.e., $Q_i = \{ a_i \in A_i \mid \sigma_i(a_i) > 0 \}$. The support of a profile $\sigma$ is the product of the supports of its individual strategies $\sigma_i$.
\end{definition}
We now define the notion of \emph{Nash equilibria} (NE), which are strategy profiles
for which there is no incentive for any player to unilaterally change their strategy.

\begin{definition}[Best response] \label{def:best}
For strategy tuple $\sigma_{-i} \in \Sigma^{-i}_\nfgame$,
a \emph{best response} to $\sigma_{-i}$ for player $i$ is a strategy $\sigma^\star_i \in \Sigma^i_\nfgame$ such that $u_i(\sigma_{-i}[\sigma^\star_i]) \geq u_i(\sigma_{-i}[\sigma_i])$ for all $\sigma_i \in \Sigma^i_\nfgame$.
\end{definition}
\begin{definition}[Nash equilibrium]\label{def:nash}
A strategy profile $\sigma^\star$ is a \emph{Nash equilibrium} (NE) and $\langle u_i(\sigma^\star)\rangle_{i \in N}$ \emph{NE values}
if $\sigma_i^\star$ is a best response to $\sigma_{-i}^\star$ for all $i \in N$.
\end{definition}
Since multiple NE can exist for an NFG, we are also interested in
finding the \emph{optimal} equilibrium for a given criterion.
In this paper, we focus on \emph{social welfare}
NE, which are those that maximise the sum of the players' utilities.
\begin{definition}[Social welfare NE]\label{def:swne}
An NE $\sigma^\star$ is a \emph{social welfare optimal NE} (SWNE) and $\langle u_i(\sigma^\star)\rangle_{i \in N}$ corresponding \emph{SWNE values} if $u_1(\sigma^\star){+}\cdots{+}u_n(\sigma^\star)\geq u_1(\sigma){+} \cdots{+}u_n(\sigma)$ for all NE $\sigma$. 
\end{definition}
We are now ready to discuss \emph{normal form psychological games} (NFPGs)~\cite{GPS89},
a generalisation of NFGs in which a player's utility
can depend not only on the game's outcome (actions taken),
but also on the player's \emph{belief} as to the outcome.
The notions of actions, strategies and strategy profiles remain the same as for NFGs
and, for NFPG $\nfpgame$, we use the same notation
$\Sigma^i_\nfpgame$, $\Sigma_\nfpgame$ and $\Sigma^{-i}_\nfpgame$.

\startpara{Beliefs and coherence}
The \emph{first-order beliefs} for player $i$ represent their belief
as to the (mixed) strategies that will be taken by the other players.
So, the set of all first-order beliefs for player $i$
is defined as $\Belief_i^1 = \Dist(\Sigma_\nfpgame^{-i})$.
\emph{Higher-order beliefs} are beliefs about the beliefs of other players.
We will denote the $k$th order beliefs for player $i$ by $\Belief_i^k$
and write $\smash{\Belief_{-i}^k = \prod_{j\neq i} \Belief_{j}^k}$
for the set of $k$th order beliefs for all players other than $i$.
Higher-order beliefs are then defined inductively by: 
\[
\Belief_i^{k+1} \ \rmdef \ \Dist(\Sigma_\nfpgame^{-i} \times \Belief_{-i}^1 \times \cdots \times \Belief_{-i}^k) \, .
\]
Notice that information about beliefs appears multiple times.
This allows for correlation between different orders of belief,
e.g., second-order beliefs $\Belief_i^2$ assign probabilities to combinations of
the strategies $\Sigma_\nfpgame^{-i}$ and first-order beliefs $\Belief_{-i}^1$ of the other players.
As in \cite{GPS89}, we will assume that beliefs are \emph{coherent},
meaning that this information is consistent.
For example, the marginal of player $i$'s second-order beliefs with respect to $\Sigma_\nfpgame^{-i}$
should coincide with $i$'s first-order beliefs.
The same condition is applied inductively to higher-order belief sets.

Furthermore, since players are rational and know that other players are also rational,
coherency is assumed to be common knowledge and we will require beliefs to be \emph{collectively coherent}.
In other words, each player $i$ only ever believes
that another player $j$'s beliefs are coherent,
that player $j$ believes other players to be coherent, and so on.
We will write $\Belief_i$ for the set for all collectively coherent higher-order beliefs
for player $i$, and define $\Belief = \prod_{i\in N} \Belief_i$
to be the set of collectively coherent \emph{belief profiles},
i.e., the set of beliefs for all players.

\startpara{Psychological games}
We can now formally define the psychological variant of normal form  games, where the key difference is that the utility for player $i$
now also depends on their (collectively coherent) belief $b_i\in\Belief_i$
about the other players, as well as the actions they actually take.
 
\begin{definition}[Normal form psychological game] A (finite, $n$-person) \emph{normal form psychological game} (NFPG) is a tuple $\nfpgame = (N,A,u)$ where:
\begin{itemize}
    \item $N=\{1,\dots,n\}$ is a finite set of players;
    \item $A_i$ is a finite set of \emph{actions} available to player $i \in N$ and $A_i \cap A_j = \emptyset$ for $i \neq j \in N$;
    \item $u = (u_1,\dots,u_n)$ and $u_i \colon ( \Belief_i \times A ) \rightarrow \Qset$
    is a utility function for player $i \in N$.
\end{itemize}
\end{definition}
As for NFGs, we can define the \emph{expected} utility of player $i$ for a given
strategy profile $\sigma$.
However, here we must now also include beliefs. More precisely, for
belief $b_i\in\Belief_i$ for player $i$ and strategy profile~$\strat$,
we write $u_i(b_i,\strat)$ for the expected utility of player $i$ under $b_i$ and $\strat$.

We can now define the notion of \emph{psychological Nash equilibrium} (PE).
While an NE for an NFG $\nfgame$ is a strategy profile $\sigma\in\Sigma_{\nfgame}$,
a PE for an NFPG $\nfpgame$ is a pair $(b,\sigma)\in\Belief\times\Sigma_{\nfpgame}$
comprising a belief profile $b$ and a strategy profile $\sigma$.
Crucially, as explained in~\cite{GPS89}, it is assumed that, in equilibrium,
the beliefs $b$ of the players match a commonly held view of reality.
In other words, each player $i$ believes, with probability 1,
that each other player $j$ follows strategy $\sigma_j$,
that player $j$'s beliefs match $\sigma_{-j}$, and so on.
For strategy profile $\sigma$, this matching belief profile
for all $n$ players is denoted $\beta(\sigma)$.

\begin{definition}[Psychological Nash equilibrium]\label{def:pe}
A pair $(b^\star,\sigma^\star)$ 
of belief profile $b^\star=(b^\star_1,\dots,b^\star_n)\in\Belief$
and strategy profile $\sigma^\star=(\sigma^\star_1,\dots,\sigma^\star_n)\in\Sigma_{\nfpgame}$
for NFPG $\nfpgame$ is a \emph{psychological Nash equilibrium} (PE)
if: 
\begin{align}
    &b^\star=\beta(\sigma^\star) \label{pe1new-eqn} \\
    &u_i(b^\star_i,\sigma^\star) \geq u_i(b^\star_i,\sigma_{-i}[\sigma_i]) \text{ for all } \sigma_i \in \Sigma^i_\nfpgame \text{ and } i\in N. \label{pe2-eqn} 
\end{align}
\end{definition}
The first condition \eqnref{pe1new-eqn} implies that, as discussed above, the players' beliefs match a commonly held view of reality. The second condition \eqnref{pe2-eqn} matches the corresponding requirements for NEs of NFGs
(see Definitions~\ref{def:best} and \ref{def:nash}).
As for NEs, we will generally aim to find a PE that is \emph{social welfare optimal},
where the sum of the players' utilities is maximised.

\startpara{Defining utility functions}
Since we focus on psychological Nash equilibria, condition~\eqnref{pe1new-eqn} above,
combined with the assumption of collective coherence for higher-order beliefs,
allows us to adopt a simpler formulation of an NFPG's utility functions in practice.
Although player $i$'s utility function $u_i$ depends on its
(collectively coherent, higher-order) beliefs about the other players, since we 
know that there is a common belief in equilibrium we can simply define $u_i$ in terms of the strategies alone, that is, as a function of the probabilities that each player $j$ takes each of its actions. Additionally, for simplicity,
we allow each player's utility to be defined
in terms of \emph{all} the player's strategies,
including the players' \emph{own} choices of actions.

\begin{examp}\label{ex:confidence}
To illustrate NFPGs, let us consider the \emph{confidence} game from~\cite{GPS89}, which comprises three players. Player 1 submits a proposal, which is randomly assigned with equal probability to player 2 or 3.
They can then chose to either $\mathit{accept}$ or $\mathit{reject}$ this proposal.
We abbreviate these actions to $a_i$ and $r_i$, respectively, for player $i=2,3$.
Player 1 has no actions to take.

The game has belief-dependent utilities,
involving both first-order and second-order beliefs. We will write $\probact{a}$ for the \emph{probability} that a player $i$ chooses action $a$ in their (mixed) strategy,
$\overline{\probact{a}}$ for the expectation of another player $j \neq i$
as to the probability~$\probact{a}$, i.e., the first-order belief for player $j$,
and $\doverline{\probact{a}}$ for the expected value of $\overline{\probact{a}}$
from the perspective of another player, i.e., the second-order belief.

Player 1 wants the proposal to be accepted, but their satisfaction about acceptance is influenced by their belief about how likely this is to happen: being more \emph{optimistic} means they are happier about an acceptance, but also much unhappier in the case of a rejection.
Player 2 is influenced by how \emph{confident} player 1 is about acceptance,
and is more likely to accept the proposal if they believe player 1 is more confident.
Player 3 always prefers rejection.

These notions are encoded in the players' utility functions as follows,
where we follow~\cite{GPS89} but show extra details of the derivation.
Let us denote the probability that player 1's proposal is accepted as $\probact{acc}$.
We have $\probact{acc}=(1/2){\cdot}(\probact{a_2}+\probact{a_3})$
since players 2 and 3 are assigned the proposal with equal probability.
Similarly, $\overline{\probact{acc}}=(1/2){\cdot}(\overline{\probact{a_2}}+\overline{\probact{a_3}})$
is player 1's belief as to the likelihood of acceptance,
and $\doverline{\probact{acc}}=(1/2){\cdot}(\doverline{\probact{a_2}}+\doverline{\probact{a_3}})$
equals player 2's belief about $\overline{\probact{acc}}$.

Player 1 has a utility of 1 in case of acceptance, plus a further utility based on their degree of optimism, i.e., in terms of belief of acceptance, of $2\cdot\overline{\probact{acc}}$. Conversely, if rejected, their utility is $-8\cdot\overline{\probact{acc}}$.
Player 2 prefers to accept when $\doverline{\probact{acc}}>\frac{1}{6}$ so,
when assigned the proposal, receives a utility of $6\cdot\doverline{\probact{acc}}$ for accepting and 1 for rejecting.
Player 3 has a utility of 1 if they are assigned the proposal and reject it, otherwise 0.

Recall from above that, in equilibrium,
players' beliefs must match a shared view of reality,
so $\probact{a_i}=\overline{\probact{a_i}}=\doverline{\probact{a_i}}$ for $i=2,3$. We therefore express the utility functions for players as expressions in terms
of just $\probact{a_i}$.
Since player 1 does not choose an action and players 2 and 3 each can choose between two actions,
we can write player $i$'s utility function as a $2{\times}2$ matrix~$\mgame_i$.
For each pair of actions of players 2 and 3, the value combines the utility arising when each of
player 2 and 3 are assigned the proposal, weighted by probability 1/2:

\begin{align*}
\mgame_1 & = 
\bordermatrix{
 & {a_3} & {r_3} \cr 
 {a_2} & 1+\probact{a_2}+\probact{a_3} & 1/2 - (3/2){\cdot}(\probact{a_2}+\probact{a_3}) \cr 
 {r_2} & 1/2 - (3/2){\cdot}(\probact{a_2}+\probact{a_3}) & -4{\cdot}(\probact{a_2}+\probact{a_3})
 }
\\
\mgame_2 & = 
\bordermatrix{
 & {a_3} & {r_3} \cr 
 {a_2} & (3/2){\cdot}(\probact{a_2}{+}\probact{a_3}) & (3/2){\cdot}(\probact{a_2}{+}\probact{a_3})  \cr 
 {r_2} & 1/2 & 1/2 
 }
\qquad
\mgame_3 = 
\bordermatrix{
 & {a_3} & {r_3} \cr 
 {a_2} & 0 & 1/2 \cr 
 {r_2} & 0 & 1/2
 }
\end{align*}

\noindent
When determining PE for the confidence game,
we can ignore player 1 since it has no actions. In an equilibrium, suboptimal actions cannot be played with positive probability.
We can therefore compute a solution by encoding the problem with the following set of constraints:
\begin{align}
 \big((3/2){\cdot}((\probact{a_2} + \probact{a_3}){\cdot} \probact{a_3} + (\probact{a_2} + \probact{a_3}){\cdot} \probact{r_3}) \geq (1/2){\cdot} (\probact{a_3} + \probact{r_3}) \big) \lor (\probact{a_2}=0) \label{eg1-eqn} \\
 \big((1/2){\cdot}(\probact{a_3} + \probact{r_3}) \geq (3/2){\cdot}((\probact{a_2} + \probact{a_3}){\cdot} \probact{a_3} + (\probact{a_2} + \probact{a_3}){\cdot} \probact{r_3}) \big) \lor (\probact{r_2}=0) \label{eg2-eqn} \\
 \big(0 \geq (1/2){\cdot}(\probact{a_2} + \probact{r_2}) \big) \lor (\probact{a_3} = 0)  \label{eg3-eqn} \\
 \big((1/2){\cdot}(\probact{a_2} + \probact{r_2}) \geq 0 \big) \lor (\probact{r_3} = 0)    \label{eg4-eqn}
\end{align}
For example, \eqnref{eg1-eqn} must hold because either the action $a_2$ is optimal for player 2, i.e., the utility obtained by player 2 when action $a_2$ is chosen is greater than or equal to that when action $r_2$ is chosen under the optimal strategy of player 3, or the action $a_2$ is chosen with probability 0.
Any assignment that satisfies all four constraints is an equilibrium. Given that $\probact{a_2}+\probact{r_2} = 1$, the first clause of \eqnref{eg3-eqn} cannot be satisfied and thus we must have $\probact{a_3} = 0$. This is consistent with the fact that $a_3$ is dominated for player 3, i.e., action $r_3$ always yields higher utility than $a_3$. If $\probact{a_3} = 0$, we have $\probact{r_3} = 1$, which means that the second clause of \eqnref{eg4-eqn} has to be false. The first clause of \eqnref{eg4-eqn} is trivially satisfied. Constraints \eqnref{eg1-eqn} and \eqnref{eg2-eqn} can then be reduced to:
\begin{align}
 (3{\cdot}\probact{a_2} \geq 1) \lor (\probact{a_2}=0)\\ 
 (1 \geq 3{\cdot}\probact{a_2}) \lor (\probact{r_2}=0)
\end{align}
We then obtain satisfying assignments by setting $\probact{a_2}=1/3$ and $\probact{r_2}=2/3$, $\probact{a_2}=1$ and $\probact{r_2}=0$ or $\probact{a_2}=0$ and $\probact{r_2}=1$ with utility vectors $u=(u_1,u_2,u_3)$ equal to $(-8/9,1/2,1/2)$, $(-1,3/2,1/2)$ and $(0,1/2,1/2)$, respectively. The proposal has the highest chance of being accepted in the second equilibrium, when player 1 is most confident. However, as player 3 is certain to reject, that is also the worst equilibrium for player 1, who is bound to be disappointed. The last two  equilibria are social welfare optimal with a combined utility of 1.
\end{examp}

%% file: modelchecking.tex
\section{Equilibria Computation for Psychological Games}\label{sec:equcomp}

We now propose methods for analysing NFPGs in order to
determine their PEs and corresponding values. The approach builds upon techniques for the non-psychological setting, i.e., finding NEs for NFGs. For the case of two-player NFGs (bimatrix games), we can use well known approaches such as the Lemke-Howson \cite{LH64} algorithm or mixed-integer programming based on regret minimisation~\cite{SGC05}. For NFGs with more than two players, algorithms include the Govindan-Wilson~\cite{SW03} or Simplicial Subdivision~\cite{LTH87}, as well as search methods based on \emph{support enumeration}~\cite{PNS04}.

We take the support enumeration approach for NFPGs, by adapting the method of~\cite{KNPS20b},
which has 
been used to find social welfare optimal NEs for NFGs in a similar fashion.
This approach exhaustively inspects sub-regions of the strategy profile space,
based on the idea that searching for NEs within a specific \emph{support} (see \defref{def:support}) of a strategy profile is computationally easier. It relies on encoding the computation
of a (social welfare optimal) NE as a \emph{non-linear programming} (NLP) problem.

The NLP encoding leverages conditions for a strategy profile to characterise an NE, presented as a lemma in~\cite{KNPS20b}, and based on the notion of \emph{feasibility program} introduced in \cite{DK93,PNS04}. 
The lemma states that a strategy profile of an NFG is an NE if and only if any player switching to a single action in the support of the profile yields the same utility for the player,
and switching to an action outside the support can only decrease its utility. We adapt that lemma here to characterise a PE of an NFPG. This result follows directly from Definition~\ref{def:pe} and, in particular, the fact that in equilibrium the belief profile needs to correspond to the strategies being played.
\begin{lemma}\label{lem:nash} A pair $(b,\sigma)$ comprising a belief profile $b$ and a strategy profile $\sigma{=}(\sigma_1,\dots,\sigma_n)$ of $\nfpgame = (N,A,u)$ is a PE if and only if \eqnref{pe1new-eqn} and the following conditions are satisfied: 
\begin{eqnarray}
\forall i \in N . \, \forall a_i \in A_i . \, && \sigma_i(a_i)>0 \rightarrow u_i(b,\sigma_{-i}[\eta_{a_i}]) = u_i(b,\sigma) \label{necond1-eqn} \\
\forall i \in N . \, \forall a_i \in A_i . \, && \sigma_i(a_i)=0 \rightarrow u_i(b,\sigma_{-i}[\eta_{a_i}]) \leq u_i(b,\sigma) \,  \label{necond2-eqn}
\end{eqnarray}
where $\eta_{a_i}$ is the pure strategy that picks action $a_i$ with probability 1.
\end{lemma} 

We now extend the NLP 
encoding of the computation of an NE presented in \cite{KNPS20b} to the case of an NFPG $\nfpgame = (N,A,u)$.
Since this encoding uses a support enumeration approach,
we need to determine the social welfare optimal PE
amongst strategy profiles from a fixed support $Q = Q_1 {\times} \cdots {\times}Q_n \subseteq A$, i.e., for a given set of actions of each player.
We first choose a \emph{pivot} action $q^p_i \in Q_i$, which can be any action in $Q_i$, for each player $i$.
The problem is then to maximise:
\begin{equation}\label{opt-eqn}
\begin{array}{c}
\sum_{i \in N} \left( \sum_{q \in Q} u_i(b^\star,q) \cdot \left( \prod_{j \in N} p_{q_j} \right) \right)
\end{array}
\end{equation}
subject to:
\begin{eqnarray}
\!\!\!\!\!\!\!\!\!\!\mbox{$\sum\limits_{c \in Q_{-i}(q_i^p)} u_i(b^\star,c) \cdot \left( \prod\limits_{j\in N_{-i}} p_{c_j} \right)
- \sum\limits_{c \in Q_{-i}(q_i)} u_i(b^\star,c) \cdot \left( \prod\limits_{j\in N_{-i}} p_{c_j} \right)$} & = & 0 \label{eq-eqn} \\
\!\!\!\!\!\!\!\!\!\!\mbox{$\sum\limits_{c \in Q_{-i}(q^p_i)} u_i(b^\star,c) \cdot \left( \prod\limits_{j\in N_{-i}} p_{c_j} \right)
- \sum\limits_{c \in Q_{-i}(a_i)} u_i(b^\star,c) \cdot \left( \prod\limits_{j\in N_{-i}} p_{c_j} \right)$} & \geq & 0  \label{leq-eqn} \\
\mbox{$\sum\limits_{q_i \in Q_i} p_{q_i}$} =  1 \quad \mbox{and} \quad p_{q_i} &  >  &  0 \label{ge-eqn}
\end{eqnarray}
for all $i \in N$, $q_i \in Q_i {\setminus} \{q_i^p \}$ and $a_i \in A_i {\setminus} Q_i$ where $Q_{-i}(c_i) = Q_1 {\times} \cdots {\times} Q_{i-1} {\times} \{ c_i \}$ ${\times} Q_{i+1} {\times} \cdots {\times} Q_n$, $N_{-i} = N {\setminus} \{i\}$ and $b^\star \in \Belief$.

The variables $p_{q_i}$ represent the probabilities of players choosing different actions, i.e. the probability player $i$ selects action $q_i \in Q_i$. If a satisfying assignment is found, we have a social welfare optimal PE given by the belief and strategy profiles pair $(b^\star, \sigma^\star)$,  where, for $a_i \in A_i$, $\sigma_i^\star(a_i)=p_{q_i}$ if $a_i=q_i$ and $q_i \in Q_i$, and 0 otherwise. Following condition \eqnref{pe1new-eqn} of Definition~\ref{def:pe}, we have $b^\star=\beta(\sigma^\star)$.

Constraints \eqnref{eq-eqn} and \eqnref{leq-eqn} enforce that the solution corresponds to a PE, encoding constraints \eqnref{necond1-eqn} and \eqnref{necond2-eqn}, respectively, of \lemref{lem:nash} when restricting to pivot actions. This restriction is sufficient as \eqnref{necond1-eqn} requires all actions in the support to yield the same utility. The objective function in \eqnref{opt-eqn} corresponds to the sum of the individual utilities of the players when they play according to the profile corresponding to the solution. By maximising it, we require the solution to be social welfare optimal. As it is possible to have more than one equilibrium for which the sum of utilities is optimal, we specify additional \emph{lower priority} objectives to maximise individual payoffs following an increasing sequence of indices $i$, and thus output a payoff vector with a consistent ordering. 

\begin{figure}[t]
\raisebox{1.25cm}{
\begin{subfigure}{0.4\textwidth}
\begin{align*}
&\mgame_{1} = 
\bordermatrix{
 & a_2 & b_2 \cr 
 \bot & -\frac{400}{81}{\cdot}\probact{a_2}{+}\frac{40}{9} & 0
}
\\
&\mgame_{2} = 
\bordermatrix{
 & a_2 & b_2 \cr 
 \bot & 0 & 0 
}
\end{align*}
\end{subfigure}
}
\hfil
\begin{subfigure}{0.7\textwidth}
    \centering
    \begin{tikzpicture}
    \begin{axis}[
        xlabel = {$\probact{a_{\scale{.75}{2}}}$},
        ylabel = {\scriptsize \emph{utility}},
        y label style={at={(axis description cs:0.1,.5)},anchor=south
        },
        width=0.7\textwidth,
        height=0.5\textwidth,
        xmin=0.0, xmax=1.0,
        ymin=-0.5, ymax=1,
        ytick={-0.5,-0.25,0,0.25,0.5,0.75,1.0},
        domain=0:2,
    ]
    \addplot[teal,mark=o] {-(400/81)*x^2+(40/9)*x};
    \end{axis}
    \end{tikzpicture}
\end{subfigure}
\vspace*{-2em}
\caption{Player utilities (left) and total expected utility (right) for Example~\ref{examp:solve}.}
\label{fig:ex_swne}
\end{figure}
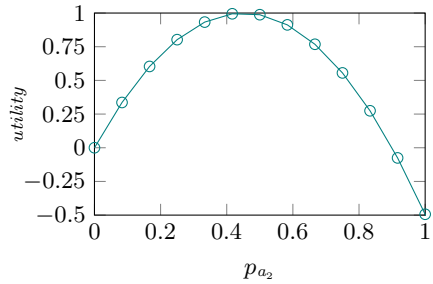

\begin{examp}\label{examp:solve}
Consider the two-player NFPG whose utility functions are given by the matrices in Figure~\ref{fig:ex_swne} (left).
Player 1 has no choice (we write $A_1=\{\bot\}$) and player 2 chooses an action from $A_2=\{a_2,b_2\}$.
Player 2 is indifferent (their utility is always 0), whereas player 1's utility depends on
(their expectation about) the probability $\probact{a_2}$ of $a_2$ being played.
Therefore, player 1's expected utility (which is also the total expected utility)
is a function depending only on $\probact{a_2}$:
\[
    u_1(\probact{a_2}) = -\frac{400}{81}{\cdot}\probact{a_2}^{\!2}+\frac{40}{9}{\cdot}\probact{a_2}
\]
Since the only player with a choice is indifferent between their own actions, the constraints in \eqnref{eq-eqn} and \eqnref{leq-eqn} are trivially satisfied for all supports as long as \eqnref{ge-eqn} is also satisfied, and thus any strategy profile (with an accompanying belief profile) is an equilibrium. Figure~\ref{fig:ex_swne} (right) plots the total expected utility.
This shows that, in order to achieve the maximum value of 1,
player 2 has to randomise, picking actions $a_2$ and $b_2$ with probabilities $0.45$ and $0.55$, respectively.
\end{examp}
\noindent
The above example illustrates a contrast with (non-psychological) NFGs, and gives an indication of why computing optimal equilibria for NFPGs is more computationally challenging. For NFGs, it suffices to consider pure strategies only when finding optimal equilibria with a single active player. This can be exploited~\cite{KNPS21}, avoiding the need to solve an optimisation problem for a given support. For NFPGs, the non-linearity in the function for expected utility means that pure strategies no longer suffice for an indifferent player.

%% file: csgs.tex
\section{Psychological Concurrent Stochastic Games}\label{sec:csgs}

We next consider concurrent stochastic games (CSGs)~\cite{Sha53}, which are multi-stage games played over graphs where, at each state, players make simultaneous choices that cause the game's state to be probabilistically updated.
We present a \emph{psychological} variant of CSGs, in which, similarly to NFPGs, a player's utility (reward accumulated over a finite horizon) can depend on its \emph{belief} as to the strategies to be played as well as the actions that players select. 
We outline a procedure to compute equilibria for a class of such games, which restricts the nature of the players' beliefs.
As in previous work for CSGs~\cite{KNPS20}, we consider \emph{subgame perfect} equilibria,
which are equilibria at every state of a CSG.
\begin{definition}[Concurrent stochastic game] A \emph{concurrent stochastic multi-player game} (CSG) is a tuple
$\game = (N, S, \sinit, A, \Delta, \delta)$ where:
\begin{itemize}
\item $N=\{1,\dots,n\}$ is a finite set of players;
\item $S$ is a finite set of states and $\sinit \in S$ is the initial state;
\item $A = (A_1\cup\{\bot\}) {\times} \cdots {\times} (A_n\cup\{\bot\})$ where $A_i$ is a finite set of actions available to player $i \in N$, $A_i \cap A_j = \emptyset$ for $i \neq j \in N$
and $\bot$ is an idle action disjoint from the set $\cup_{i=1}^n A_i$;
\item $\Delta \colon S \rightarrow 2^{\cup_{i=1}^n A_i}$ is an action assignment function;
\item $\delta \colon (S {\times} A) \rightarrow \dist(S)$ is a probabilistic transition function.
\end{itemize}
\end{definition}
We assume without loss of generality that, for any $i,j \in N$, $i \neq j$, $A_i~\cap~A_j = \varnothing$. A given CSG $\game$ starts in the initial state $\sinit$ and, when in state $s$, each player $i \in N$ selects an action from its available actions $A_i(s) \rmdef \Delta(s) \cap A_i$ if this set is non-empty, and from $\{ \bot \}$ otherwise. Assuming each player $i$ selects action $a_i$, the next state of the game is determined according to the distribution $\delta(s,(a_1,\dots,a_n))$. A \emph{path} of $\game$ is a sequence $\pi = s_0 \xrightarrow{\alpha_0} s_1 \xrightarrow{\alpha_1} \cdots$ where $s_k \in S$, $\alpha_k = (a^k_1,\dots,a^k_n) \in A$, $a^k_i \in A_i(s_k)$ for $i \in N$ and $\delta(s_k,\alpha_k)(s_{k+1})>0$ for all $k \geq 0$. We denote by $\pi(i)$ the $(i{+}1)$th state of $\pi$, $\pi[i]$ the action associated with the $(i{+}1)$th transition and, if $\pi$ is finite, $\last(\pi)$ the final state. Let $\fpaths$ and $\ipaths$ denote the sets of finite and infinite paths that start in the initial state, respectively.

A \emph{strategy} for a player in $\game$ resolves the player's choices in each state. These choices can depend on the execution history and can be randomised, i.e., are of the form $\sigma_i : \fpaths \ra \dist(A_i)$ such that if $\sigma_i(\pi)(a_i) >0$, then $a_i \in A_i(\last(\pi))$. As for NFGs, a \emph{strategy profile} for $\game$ is a tuple $\sigma = (\sigma_1,\dots,\sigma_{n})$ of strategies for all players. For a given strategy profile $\strat$, a probability measure $\Prob^{\sigma}_{\game}$ over the infinite paths of $\game$ can then be defined in the standard way~\cite{KSK76}.

\startpara{Psychological CSGs}
In order to introduce a psychological variant of CSGs,
we incorporate a notion of beliefs,
and then use them to define rewards.
Let $\Belief_i^A$ denote, as defined in \sectref{sect:prelims},
the set of (collectively coherent, higher-order) beliefs for player $i$,
where first-order beliefs are over the set of actions $A$.
A \emph{belief} $b_i$ for player is of the form $b_i: S \ra \Belief_i^A$.
It is \emph{state-based}, in that it provides a separate belief for each state $s$ of the CSG,
and \emph{local}, in that these beliefs give the player's
expectations regarding the actions to be played in $s$,
not about a more global notion of the player's strategy.
A \emph{belief profile} is a tuple $b=(b_1,...,b_n)$.

A \emph{reward structure} for player $i$ takes the form $r_i=(r_i^A,r_i^S)$,
where $r_i^A \colon (S \times \Belief_i^A \times A) \ra \Qset$
is an action reward function (which maps a state, belief and action tuple to a rational value that is accumulated when the action tuple is selected in the state, assuming a given local belief for player $i$ in that state) and $r_i^S \colon S \ra \Qset$ is a state reward function (which maps each state to a rational value that is accumulated when the state is passed through).

The \emph{utility} (or \emph{objective}) for a player $i$ in CSG $\game$ can be defined by a random variable $X_i : \ipaths \to \Rset$ mapping infinite paths to reals. We denote by $\Eset^{\sigma}_{\game}(X_i)$ the expected value of player $i$'s utility under $\sigma$, with respect to the probability measure $\Prob^{\sigma}_{\game}$. Given utilities $X_1,\dots,X_n$ for all the players of $\game$, we can then define (social welfare) psychological Nash equilibria, as for NFPGs.
We will restrict our attention to utilities that correspond to \emph{finite-horizon} objectives, which may be used to investigate, for instance, the expected reward accumulated over $k$ steps. Such utilities can be expressed by a finite bound $k \in \Nset$ and reward structure $r_i=(r_i^A,r_i^S)$, with corresponding random variable:
\[X_i(\pi) = \mbox{$\sum_{j=0}^{k-1}$} \big( r_i^A(\pi(j),b_i(\pi(j)),\pi[j])+r_i^S(\pi(j)) \big) \, .\]

\startpara{Psychological equilibria computation}
In both \cite{GPS89} and \cite{BD09}, the authors point out the limitations of applying backward induction to computing equilibria for extensive (multi-stage) psychological games and show why that approach cannot be applied to the general case. While a full discussion is outside the scope of this paper, it suffices to imagine a game in which a player's utility in a given state depends on the beliefs they have about actions performed in the \emph{preceding} state. A backward induction algorithm \emph{can}, however, be applied under the assumption that psychological utilities at any given state have to be over \emph{local} strategies, that is, concerning the actions taken at that same state and coherent with an equilibrium solution for the NFPG in that state. We note that a more general approach has been proposed in \cite{BD09}, which we leave as future work.  

Using the above restriction, we devise a backward induction algorithm that builds, at each iteration, an NFPG for every state $s$ in $\game$ according to the reward structure in $s$ and the values computed for its successors in the previous iteration (initially 0 for all states). We then compute equilibria values and strategies by solving the corresponding NLPs following the definitions in Section~\ref{sec:equcomp}. We compute equilibria which are \emph{locally} social welfare optimal. Other criteria, e.g., \emph{social cost} \cite{KNPS21} or \emph{social fair} \cite{KNPS22} equilibria, which minimise the overall sum or the difference between the highest and lowest utilities, respectively, could also be applied. In the latter case, however, additional constraints would have to be added to the NLP in Section~\ref{sec:equcomp}, which would significantly increase the complexity of the problem due to the need to numerically encode logical implications. Finding all equilibria for a CSG is generally intractable, as the number may be exponential even with respect to the size of the normal form game at each state.

%% file: experiments.tex
\section{Case Studies and Experimental Results}\label{sect:expts}

We have built a prototype implementation to model and solve psychological games, and used it to investigate the applicability and performance of our approach on a selection of normal form and multi-stage psychological games.
We first consider two-player instances of the \emph{ultimatum} and \emph{reciprocity} games of~\cite{BD22}, which exemplify how psychological games can also be used in the computation of \emph{fairness equilibria}, as well as how psychological utilities can influence the strategies of the players. We then present 
two- and multi-player normal form games modelling \emph{traffic} interactions between pedestrians, cyclists and vehicles, one of which is then extended to a psychological CSG, used to  investigate how information on past decisions can influence players' strategies. 

\startpara{Implementation}
We build on top of the PRISM-games model checker~\cite{KNPS20},
extending its existing modelling language for CSGs
(in which normal form games can also be encoded as simple instances).
The key difference is that in PG models the specification of reward structures needs to incorporate a player's beliefs about the other players' strategies.
Since we currently only allow for beliefs in CSGs over local strategies (see \sectref{sec:csgs}),
rewards for a state can only make reference to (the probability of) actions played in that state.
Figure~\ref{fig:rew_example} shows a reward structure definition
in our extension of the PRISM-game modelling language
for the ultimatum game example (see \sectref{sec:rec-ult}, below). For simplicity, in this syntax, we just use the name of the action to denote the probability of choosing it, e.g., {\tt reject} denotes what we refer to elsewhere as $\probact{\mathtt{reject}}$.

As for regular CSGs, our extension of PRISM-games constructs and stores PG models using the tool's Java-based `explicit' engine. In contrast to CSGs, reward structures for PG models are represented by symbolic expressions over variables representing action choice probabilities and cannot be evaluated prior to model checking. We use Gurobi \cite{gurobi} to solve the NLPs described in \sectref{sec:equcomp} for finding equilibria values of NFPGs at each state.

\lstdefinestyle{ultimatum}{
    keywords=[1] {fair,reject,greedy,accept}
}

\begin{figure}[t]
\centering
\begin{minipage}{0.85\textwidth}
\begin{lstlisting}[language=Prism, style=ultimatum
]
// Constants
const double theta1; // Player 1's reciprocity sensitivity
const double theta2; // Player 2's reciprocity sensitivity
// Rewards for player 1
rewards "r1"
	[fair,reject] true : 5+theta1*(-4.5)*(2+0.5*reject);
	[fair,accept] true : 5+theta1*(4.5)*(2+0.5*reject);
	[greedy,reject] true : 0+theta1*(-4.5)*(-2-0.5*reject);
	[greedy,accept] true : 9+theta1*(4.5)*(-2-0.5*reject);
endrewards
// Rewards for player 2
rewards "r2"
	[fair,reject] true : 5+theta2*(-4.5)*(2+0.5*reject);
	[fair,accept] true : 5+theta2*(4.5)*(2+0.5*reject);
	[greedy,reject] true : 0+theta2*(-4.5)*(-2-0.5*reject);
	[greedy,accept] true : 1+theta2*(4.5)*(-2-0.5*reject);
endrewards
\end{lstlisting}
\end{minipage}
\vspace{-0.5em}
\caption{Reward structures for the ultimatum game, modelled in PRISM-games.}
\label{fig:rew_example}
\end{figure}

\startpara{Efficiency and Scalability} Computing equilibria values and strategies can be a complex task, even for the simpler case of finding an arbitrary (non-optimal) equilibrium of a two-player normal form game. Finding optimal equilibria of multi-player games is considerably harder, given the increased number of supports and the non-linearity of the constraints. The addition of psychological utilities complicates the computation further, as there are no natural restrictions on how these utilities may vary given the players' strategies and beliefs. At the state level, when looking for an optimal equilibrium, we are required to solve an NLP for each support. Given the total number of supports is exponential in the number of actions, i.e., equals $\prod_{i=1}^n (2^{|A_i|} -1)$, computing optimal values via enumeration can only be efficient for small games. 

\subsection{Reciprocity and Ultimatum Games}\label{sec:rec-ult}

We considered instances of the \emph{reciprocity} and \emph{ultimatum} games from \cite{BD22}, which are shown in Figure~\ref{fig:rec_ult_games}.
In each case, player 1 chooses between making a $\mathit{fair}$~($f$) or $\mathit{greedy}$ ($g$) proposal, and player 2 decides to $\mathit{reject}$ ($r$) or $\mathit{accept}$ ($a$) it. The rectangular boxes show the corresponding utilities of players 1 and 2, with player 1's utility being above that of player 2.
We present the games, as in \cite{BD22}, in extensive form,
with the players' decisions taken sequentially,
but will treat them as simultaneous moves in a single NFPG.
Otherwise, beliefs would no longer be local
since player~2's utility would depend on an earlier decision by player~1.

\input{figures/reciprocity_ultimatum/reciprocity_ultimatum_model}

In both games, player $i$ attempts to maximise the expectation of a utility function that depends on the chosen actions ($\alpha$) and belief ($b$) and has the form:
\[
u_i(\alpha,b)=\lambda_i(\alpha) + \theta_i {\cdot} \kappa_{ij}(\alpha,b){\cdot}\kappa_{ji}(\alpha,b)
\]

\noindent
where $\lambda_i$ is a \emph{material} utility function (these are the utilities shown in \figref{fig:rec_ult_games}), $\kappa_{ij}$ reflects player $i$'s \emph{kindness} to player $j$ (expected material payoff, which ranges from negative to positive) and $\theta_i\in \Rset_{\geq 0}$ is player $i$'s \emph{reciprocity sensitivity}. This type of game was originally studied in the context of \emph{fairness equilibria} \cite{Rab93}, in order to model and investigate scenarios in which agents are willing to sacrifice material utility to help or punish others depending on how they think they are being treated.

The concept of kindness was introduced in~\cite{Rab93} as a way to measure this type of feeling, and is calculated as the difference between the utility that player $i$ believes player $j$ will receive (given player $i$'s choice) minus the average of the minimum and maximum utilities player $i$ believes player $j$ could get for $i$'s other choices.
For instance, in the reciprocity game, if player 1 chooses \emph{fair}, $\kappa_{12}$ equals $5-1/2{\cdot}(5+(9{\cdot}\overline{\probact{r}} + 1{\cdot}(1-\overline{\probact{r}})) = 2-4{\cdot}\overline{\probact{r}}$.
As before, we use $\probact{r}$ for the probability of player 2 choosing action $r$ and $\overline{\probact{r}}$ for player 1's belief as to this value.

Reciprocating kindness is expressed by the matching of signs of $\kappa_{ij}$ and $\kappa_{ji}$. Thus, 
if, by adopting a particular strategy, player $i$ is perceived to be unkind to player $j$, $\kappa_{ij}$ will be negative, which will in return motivate player $j$ to be unkind to player $i$ so that the product of $\kappa_{ij}$ and $\kappa_{ji}$ is positive. A similar logic applies to when players are perceived to be kind.

As explained earlier in Example~\ref{ex:confidence},
when writing the matrices for players' utility values,
we can assume that $\overline{\probact{r}}=\probact{r}$ in equilibrium
and just express them as functions of the probability $\probact{r}$.
For the reciprocity game we thus have:
\begin{align*}
\mgame_{1}^\text{reciprocity} & = 
\bordermatrix{
 & r & a \cr 
 f & 5{+}\theta_1{\cdot}(-4){\cdot}(2{-}4{\cdot}\probact{r}) & 5{+}\theta_1{\cdot}(4){\cdot}(2{-}4{\cdot}\probact{r})  \cr 
 g & 1{+}\theta_1{\cdot}(-4){\cdot}(4{\cdot}\probact{r}{-}2) & 9{+}\theta_1{\cdot}(4){\cdot}(4{\cdot}\probact{r}{-}2)
}
\\
\mgame_{2}^\text{reciprocity} & = 
\bordermatrix{
 & r & a \cr 
 f & 5{+}\theta_2{\cdot}(-4){\cdot}(2{-}4{\cdot}\probact{r}) & 5{+}\theta_2{\cdot}(4){\cdot}(2{-}4{\cdot}\probact{r})  \cr 
 g & 9{+}\theta_2{\cdot}(-4){\cdot}(4{\cdot}\probact{r}-2) & 1{+}\theta_2{\cdot}(4){\cdot}(4{\cdot}\probact{r}{-}2)
}
\end{align*}
and for the ultimatum game:
\begin{align*}
\mgame_{1}^\text{ultimatum} & = 
\bordermatrix{
 & r & a \cr 
 f & 5{+}\theta_1{\cdot}(-9/2){\cdot}(2{+}\probact{r}/2) & 5{+}\theta_1{\cdot}(9/2){\cdot}(2+\probact{r}/2)  \cr 
 g & 0{+}\theta_1{\cdot}(-9/2){\cdot}(-2{-}\probact{r}/2) & 9{+}\theta_1{\cdot}(9/2){\cdot}(-2{-}\probact{r}/2)
}
\\
\mgame_{2}^\text{ultimatum} & = 
\bordermatrix{
 & r & a \cr 
 f & 5{+}\theta_2{\cdot}(-9/2){\cdot}(2{+}\probact{r}/2) & 5{+}\theta_2{\cdot}(9/2){\cdot}(2{+}\probact{r}/2)  \cr 
 g & 0{+}\theta_2{\cdot}(-9/2){\cdot}(-2{-}\probact{r}/2) & 1{+}\theta_2{\cdot}(9/2){\cdot}(-2{-}\probact{r}/2)
}
\end{align*}

\noindent
Figure~\ref{fig:reciprocity_ultimatum_swne} presents the strategies and utility values for SWNEs that we generated for the reciprocity and ultimatum games using different values of $\theta_1$ and $\theta_2$. Although the games are very similar in structure (there is an equal amount of material utility that can be split by the two players in different ways), it is possible to see how the reciprocity sensitivity affects their behaviours and overall utilities. For instance, in the ultimatum game, when $\theta_1=\theta_2=0$ the players are strictly concerned with their material utilities and $(\mathit{greedy}, \mathit{accept})$ is an acceptable SWNE as the sum of utilities is 10. It is possible to see though that, as $\theta_2$ increases, player 1 is less likely to play \emph{greedy} as the split becomes less fair for player 2, who could then retaliate by playing \emph{reject}.

We can also notice that, when $\theta_1=\theta_2=1$, the equilibrium for the reciprocity and ultimatum games is $(\mathit{fair}, \mathit{accept})$ which, despite leading to material utilities of ($5,5$) in both games, accounts for different overall utilities for the players. In the former, the utility for each player is equal to $u_i(\textit{fair}, \textit{accept}) = 5 + 1 {\cdot} 4 {\cdot} (2 - 4{\cdot}\probact{r}) = 13$, and for the latter it corresponds to $u_i(\textit{fair}, \textit{accept}) = 5 + 1 {\cdot} 9/2 {\cdot} (2 + \probact{r}/2) = 14$.

\input{figures/reciprocity_ultimatum/reciprocity_ultimatum_plot}

\subsection{Traffic Games}\label{sect:traffic}
    
We now report on a selection of case studies inspired by game-theoretic models of traffic and road user behaviour. We start with a simple one-shot game between a vehicle and a pedestrian in a road crossing scenario, and how their expectations can incentivise 
safe behaviour. Next we introduce a psychological variant of the Bayesian game presented in \cite{MB18}, which examined how cyclists would interact differently with autonomous and regular vehicles. Finally, we extend the road crossing scenario into a CSG and investigate the impact of combining information on past decisions with local expectations in a multi-stage, probabilistic model. 

\startpara{Pedestrian crossing}
We consider a scenario where a pedestrian is deciding whether to cross a road near oncoming traffic, illustrated in Figure~\ref{fig:pedestrian_crossing}. We assume that the car has a right of way and can $\mathit{reduce}$ ($r$) or $\mathit{maintain}$ ($m$) its speed, while the pedestrian may choose to $\mathit{cross}$ ($c$) or $\mathit{wait}$ ($w$). 
A psychological game can be constructed by including incentives set to discourage behaviour based on what they expect the other will do. We assume the pedestrian would be (illegally) jaywalking if they decided to cross, and so give them a negative reward proportional to the probability $\probact{c}$ of that action being taken (multiplied by a constant $\mu$), to model the pedestrian's fear of being caught and incurring a penalty. This parameterisation results in the following utility matrices for the vehicle and the pedestrian:
\[
\mgame_{\textit{vehicle}} = 
\bordermatrix{
 & w & c \cr 
 r & 1{-}\probact{w} & 1{+}\probact{c}  \cr 
 m & 1{+}\probact{w} & 1{-}\probact{c} 
}
\qquad
\mgame_{\textit{pedestrian}} = 
\bordermatrix{
 & w & c \cr 
 r & 1{-}\probact{r} & 1{+}\probact{r}{-}\mu{\cdot}\probact{c}  \cr 
 m & 1{+}\probact{m} & 1{-}\probact{m}{-}\mu{\cdot}\probact{c} 
}
\]

\input{figures/crossing/crossing}
\input{figures/crossing/crossing_alpha_5}

\noindent
Figure~\ref{fig:eq_cross_alpha} shows equilibria strategy profiles and utilities of this game for different values of $\mu$
(a more detailed version can be found in Figure~\ref{fig:eq_cross_alpha_appendix}, Appendix~\ref{appx}).
The colours for the points in the bottom and top plots for each value of $\mu$ serve to match the profiles and utilities of the pedestrian and the vehicle for different equilibria. While it is possible to see that there is always an equilibrium (displayed in red) in which the vehicle maintains its speed and the pedestrian waits, for smaller values of $\mu$ we also have an equilibrium strategy profile in which both agents make random choices. For example, in the equilibrium shown in blue for $\mu=2$, the vehicle randomly chooses between reducing and maintaining its speed with probabilities 3/4 and 1/4 respectively, while the pedestrian crosses or not with probability 1/2, potentially leading to unsafe behaviour. As the pedestrian's uneasiness about crossing grows, i.e., as $\mu$ increases, the strategy profiles in which they cross with positive probability disappear, with the only remaining profile for $\mu=5$ being the one in which they wait.

\startpara{Cyclist vs.\ vehicle}
Next, we model a cyclist and a vehicle, where the latter is either autonomous or driven by a human, at a road junction. A similar scenario was considered in~\cite{MB18}, but modelled as a Bayesian game to investigate how increasing the share of autonomous vehicles affects the rate of collisions. Figure~\ref{fig:cyclist_vehicle} shows the game in extensive form. The actions for the cyclist are $\mathit{yield}$ ($y$), $\mathit{walk}$ ($w$) and $\mathit{cycle}$ ($c$),
and the actions for the vehicle are $\mathit{go}$ ($g$) and $\mathit{stop}$ ($s$).

\input{figures/cyclist_vehicle/cyclist_vehicle_model}

\begin{table}[t]
\centering
\begin{tabular}{cccc}
$\alpha$                                 & $u_{\textit{nature}}$                  & $u_{\textit{cyclist}}$                  & $u_{\textit{vehicle}}$                  \\ \hline
\multicolumn{1}{|c|}{$(a, y, g)$} & \multicolumn{1}{c|}{0} & \multicolumn{1}{c|}{$5{\cdot}\probact{a}$} & \multicolumn{1}{c|}{7} \\ \hline
\multicolumn{1}{|c|}{$(a, y, s)$} & \multicolumn{1}{c|}{0} & \multicolumn{1}{c|}{$3{\cdot}\probact{a}$} & \multicolumn{1}{c|}{10} \\ \hline
\multicolumn{1}{|c|}{$(a, w, g)$} & \multicolumn{1}{c|}{0} & \multicolumn{1}{c|}{$-400{\cdot}\probact{a}$} & \multicolumn{1}{c|}{-500} \\ \hline
\multicolumn{1}{|c|}{$(a, w, s)$} & \multicolumn{1}{c|}{0} & \multicolumn{1}{c|}{$15{\cdot}\probact{a}$} & \multicolumn{1}{c|}{-15} \\ \hline
\multicolumn{1}{|c|}{$(a, c, g)$} & \multicolumn{1}{c|}{0} & \multicolumn{1}{c|}{$-500{\cdot}\probact{a}$} & \multicolumn{1}{c|}{-300} \\ \hline
\multicolumn{1}{|c|}{$(a, c, s)$} & \multicolumn{1}{c|}{0} & \multicolumn{1}{c|}{$20{\cdot}\probact{a}$} & \multicolumn{1}{c|}{15} \\ \hline
\end{tabular}
\quad\quad\quad
\begin{tabular}{cccc}
$\alpha$                                 & $u_{\textit{nature}}$                  & $u_{\textit{cyclist}}$                  & $u_{\textit{vehicle}}$                  \\ \hline
\multicolumn{1}{|c|}{$(h, y, g)$} & \multicolumn{1}{c|}{0} & \multicolumn{1}{c|}{$8{\cdot}\probact{h}$} & \multicolumn{1}{c|}{15} \\ \hline
\multicolumn{1}{|c|}{$(h, y, s)$} & \multicolumn{1}{c|}{0} & \multicolumn{1}{c|}{$6{\cdot}\probact{h}$} & \multicolumn{1}{c|}{1} \\ \hline
\multicolumn{1}{|c|}{$(h, w, g)$} & \multicolumn{1}{c|}{0} & \multicolumn{1}{c|}{$-400{\cdot}\probact{h}$} & \multicolumn{1}{c|}{-400} \\ \hline
\multicolumn{1}{|c|}{$(h, w, s)$} & \multicolumn{1}{c|}{0} & \multicolumn{1}{c|}{$15{\cdot}\probact{h}$} & \multicolumn{1}{c|}{7} \\ \hline
\multicolumn{1}{|c|}{$(h, c, g)$} & \multicolumn{1}{c|}{0} & \multicolumn{1}{c|}{$-500{\cdot}\probact{h}$} & \multicolumn{1}{c|}{-200} \\ \hline
\multicolumn{1}{|c|}{$(h, c, s)$} & \multicolumn{1}{c|}{0} & \multicolumn{1}{c|}{$20{\cdot}\probact{h}$} & \multicolumn{1}{c|}{7} \\ \hline
\end{tabular}
\vspace{0.2cm}
\caption{Psychological cyclist vs. vehicle game in normal form. Nature chooses between \emph{autonomous vehicle} ($a$) or \emph{human driver} ($h$).}
\label{tab:psych_cyclist_vehicle}
\vspace{-.3cm}
\end{table}

\input{figures/cyclist_vehicle/cyclist_vehicle_plot}

The utilities of the players reflect preferences over potential collisions, in accordance with traffic rules and an assumption on the part of the cyclist that an autonomous vehicle (AV) would be programmed to be as \emph{risk averse} as possible, while human drivers are often (unintentionally) distracted. So, for example, the penalty for an AV not stopping while the cyclist crosses the road is higher than for a human-driven vehicle (300 vs. 200). In the original game, it is assumed that a virtual \emph{nature} player chooses the type of the vehicle according to a prior distribution. Here, we consider \emph{nature} to be an active albeit \emph{indifferent} player picking $\mathit{autonomous}$ ($a$) or $\mathit{human}$ ($h$), and we set the utilities for the cyclist to be the original utilities multiplied by their expectation of the vehicle being driven autonomously ($p_a$) or by a human ($p_h$), as detailed in Table~\ref{tab:psych_cyclist_vehicle}. 

Figure~\ref{fig:eq_cyclist_vehicle_plot} shows results for this variant
(a more detailed version is in Figure~\ref{fig:eq_cyclist_vehicle_plot_appendix} of  Appendix~\ref{appx}).
In addition to the 
equilibria computed in~\cite{MB18} (indicated in blue, orange\footnote{The utility and strategy for the cyclist are the same as for the one in magenta.}, magenta and brown), two new equilibria (indicated in green and red) are present for the psychological variant, in which 
\emph{nature} randomly chooses between an autonomous vehicle and one with a human driver with probabilities $p_a=14/17$ and $p_h=3/17$. Considering the original model from \cite{MB18} and combining the utilities of the two 
original games
(Figure~\ref{fig:cyclist_vehicle}(a) and (b))  into one normal form game by multiplying the corresponding utilities by $p$ or $1-p$, we obtain the bimatrix game as follows:
\[
\mgame_{\textit{cyclist}} = 
\bordermatrix{
 & g & s \cr 
 y & -3{\cdot}p{+}8 &-3{\cdot}p{+}6 \cr 
 w & -400 & 15 \cr
 c & -500 & 20
}
\quad
\mgame_{\textit{vehicle}} = 
\bordermatrix{
 & g & s \cr 
 y & -8{\cdot}p{+}15 & 9{\cdot}p{+}1 \cr 
 w & -100{\cdot}p{-}400 & 14{\cdot}p{+}7 \cr
 c & -100{\cdot}p{-}200 & 8{\cdot}p{+}7
}
\]
We observe that, if the cyclist decides to \emph{yield}, the vehicle would only play \emph{go} with positive probability if $-8{\cdot}p+15 \geq 9{\cdot}p+1$, which gives $p \leq 14/17$. For any value of $p$ above that threshold, the game has only one equilibrium in (\emph{cycle}, \emph{stop}), with utilities $(20,8{\cdot}p+7)$. The analysis in~\cite{MB18} suggests that the number of collisions, i.e., the outcome where the cyclist chooses \emph{cycle} and the vehicle chooses \emph{go}, drops as the proportion of AVs increases. 
Indeed, if we only have AVs circulating, the only equilibrium (magenta) is (\emph{cycle}, \emph{stop}). However, by modelling this scenario as a three-player game with nature as an active player, we see in Figure~\ref{fig:eq_cyclist_vehicle_plot} (and in more detail in Figure~\ref{fig:eq_cyclist_vehicle_plot_appendix} of  Appendix~\ref{appx}), for a mix of AVs and human-driven vehicles, where AVs correspond to approximately 82\% of the fleet, that the equilibrium strategy (green and red) for the cyclist is actually to \emph{yield}. Furthermore, the vehicles 
can follow two different strategies: one in which they would \emph{go} with probability 1 and another in which they \emph{stop} with probability approximately $0.97$ (shown in red and green, respectively). Finally, the model in~\cite{MB18} assumes that the cyclists can always differentiate between an AV and a human-driven vehicle, which is not realistic. In contrast, our model allows for the possibility of specifying psychological payoffs, meaning the actions taken by the cyclist can vary according to their beliefs about the type of the vehicle, paving the way for more sophisticated models of similar scenarios.

\startpara{Multi-stage pedestrian crossing}
Finally, we consider a probabilistic, multi-stage version of the earlier \emph{pedestrian crossing} game, modelled as a CSG with psychological utilities. The state of the CSG has a variable $j$ counting the number of times the one-shot pedestrian game has been repeated.
We also add two discrete integer-valued variables $c_r, c_w \in \{0,1,\dots,10\}$ in order to carry information forward about the actions taken by both agents. The variables $c_r$ and $c_w$ are initialised to 0 and go up (or down) by 1 when the vehicle reduces (or maintains) its speed, and the pedestrian decides to wait (or cross), respectively.

\input{figures/multistage_crossing/multistage_crossing_model}

To account for the fact that these observations can be imperfectly made by the players, $c_r$ and $c_w$ are updated probabilistically according to an \emph{attention} coefficient $\gamma \in [0,1]$, and so their values can also remain the same with probability $1-\gamma$, as illustrated in Figure~\ref{fig:pedestrian_transitions}. The information on past decisions carried by the values of the two variables is then weighted with the local expectations at each state by modifying the reward matrices in following manner:
\begin{align*}
\mgame_{\textit{vehicle}} & = 
\bordermatrix{
 & w & c \cr 
 r & 1{-(}1/2){\cdot}(\probact{w}{+}c_w/10) & (3/2){+}(1/2){\cdot}(\probact{c}{-}c_w/10)  \cr 
 m & 1{+}(1/2){\cdot}(\probact{w}{+}c_w/10) & (1/2){-}(1/2){\cdot}(\probact{c}{-}c_w/10)  \cr 
}
\\
\mgame_{\textit{pedestrian}} & = 
\bordermatrix{
 & w & c \cr 
 r & 1{-}(1/2){\cdot}(\probact{r}{+}c_r/10) & 1{+}(1/2){\cdot}(\probact{r}{+}c_r/10){-} \mu{\cdot}\probact{c}  \cr 
 m & (3/2){+}(1/2){\cdot}(\probact{m}{-}c_r/10) & (1/2){-}(1/2){\cdot}(\probact{m}{-}c_r/10){-} \mu{\cdot}\probact{c} 
}
\end{align*}

Figure~\ref{fig:crossing_multi} shows expected utility values and crossing probability averages for different values of $\gamma$, paths of length $k$ and $\mu=1$. In a similar fashion to the one-shot example, we do not focus exclusively on the social welfare solution and there could be multiple equilibria for the NFPGs built at each state. However, for extensive or multi-stage games, it is impractical to compute all possible equilibria unless the number of states is fairly small. This is particularly true when there is probabilistic branching, as the number of equilibria (and hence the number of NLPs to be solved) may grow very rapidly. For this reason, the probability values reported in Figure~\ref{fig:crossing_multi} (right) are averages over equilibria selected uniformly at random at each state. For each path length and $\gamma$, 10 experiments were run and the utility averages for the initial state are displayed in Figure~\ref{fig:crossing_multi} (left).

\input{figures/multistage_crossing/multistage_crossing_plot}

As the value of $\gamma$ grows, it is possible to notice a trend in which the pedestrian's behaviour is safer, with the average probability of a crossing decreasing (top right). This is also reflected in higher utility for the pedestrian (top left), particularly when considering longer paths. For the vehicle, the utility averages (bottom left) remain fairly stable but we can also see that the likelihood of it reducing its speed (bottom right) decreases. This behaviour is desirable, considering it happens in coordination with decreasing probability values of a pedestrian crossing and the fact that the vehicle should have a right of way. 

To provide an indication of model sizes that we are able to analyse, Table~\ref{tab:stats} shows how the number of CSG states and transitions, and the average computation time, vary for different values of $k$ and intervals of $\gamma$ (note that the number of transitions and states is the same for any value of $\gamma \in (0,1)$, see Figure~\ref{fig:pedestrian_transitions}).

\input{table_multistage_crossing_stats}

%% file: figures/reciprocity_ultimatum/reciprocity_ultimatum_model.tex
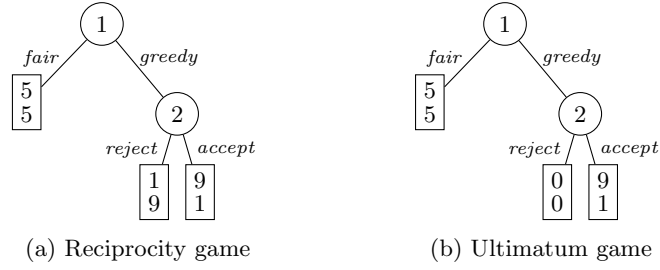
\begin{figure}[t]
\centering
    \begin{subfigure}{0.3\textwidth}
    \centering
    \begin{forest}
      [$1$, node options={draw,circle}, 
    	[{\shortstack[c]{5 \\ 5}}, node options={draw,rectangle}, xshift=-1.0cm, name=c1c1, edge={thin}, edge label={node[pos=0.4,left]{\scriptsize \emph{fair}}}]
      	[$2$, node options={draw,circle}, xshift=1.0cm, edge = {thin}, edge label={node[pos=0.3,right]{\scriptsize \emph{greedy}}}, name=p2_2, 
    		[{\shortstack[c]{1 \\ 9}}, node options={draw,rectangle}, xshift=1.0cm, name=d1c2, edge={thin}, edge label={node[midway,left,left]{\scriptsize \emph{reject}}}]
    		[{\shortstack[c]{9 \\ 1}}, node options={draw,rectangle}, xshift=1.0cm, name=d1d2, edge={thin}, edge label={node[midway,right]{\scriptsize \emph{accept}}}]
    	]
      ]
    \end{forest}
    \subcaption{Reciprocity game}
    \end{subfigure}
    \hfil
    \begin{subfigure}{0.3\textwidth}
    \centering
    \begin{forest}
      [$1$, node options={draw,circle}, 
    	[{\shortstack[c]{5 \\ 5}}, node options={draw,rectangle}, xshift=-1.0cm, name=c1c1, edge={thin}, edge label={node[pos=0.4,left]{\scriptsize \emph{fair}}}]
      	[$2$, node options={draw,circle}, xshift=1.0cm, edge={thin}, edge label={node[pos=0.3,right]{\scriptsize \emph{greedy}}}, name=p2_2, 
    		[{\shortstack[c]{0 \\ 0}}, node options={draw,rectangle}, xshift=1.0cm, name=d1c2, edge={thin}, edge label={node[midway,left]{\scriptsize \emph{reject}}}]
    		[{\shortstack[c]{9 \\ 1}}, node options={draw,rectangle}, xshift=1.0cm, name=d1d2, edge={thin}, edge label={node[midway,right]{\scriptsize \emph{accept}}}]
    	]
      ]
    \end{forest}
    \subcaption{Ultimatum game}
    \end{subfigure}
\caption{Reciprocity and ultimatum games in extensive form. The utilities for players 1 and 2 are given in the top and bottom rows of each leaf node, respectively.}
\label{fig:rec_ult_games}
\end{figure}

%% file: figures/reciprocity_ultimatum/reciprocity_ultimatum_plot.tex
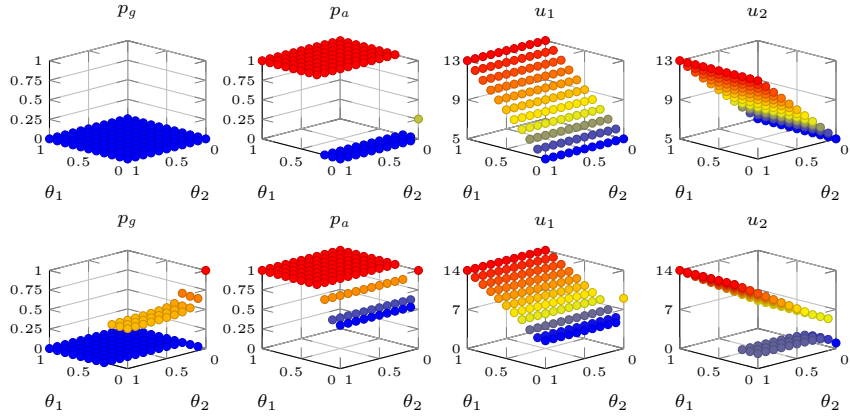
\begin{figure}[t]
    \tiny
    \centering
    \begin{subfigure}{0.3\textwidth}
        \centering
        \begin{tikzpicture}
        \begin{axis}[
            title={\scriptsize $\probact{g}$},
            width=\textwidth,
            grid=major,
            view={-135}{20},
            zmin=0.0,
            zmax=1.0,
            ztick={0,0.25,0.5,0.75,1.0},
            xlabel={\scriptsize $\theta_1$},
            ylabel={\scriptsize $\theta_2$},
            ] 
            \addplot3+[mesh,scatter,only marks,mark size=1.5pt] table[x=theta1, y=theta2, z=r, col sep=comma] {csv/reciprocity_ultimatum/reciprocity_swne_filtered.txt};
        \end{axis}
        \end{tikzpicture}        
    \end{subfigure}
    \hspace{-1.0cm}
    \begin{subfigure}{0.3\textwidth}
        \centering    
        \begin{tikzpicture}
        \begin{axis}[
            title={\scriptsize $\probact{a}$},
            width=\textwidth,
            grid=major,
            view={-135}{20},
            zmin=0.0,
            zmax=1.0,
            ztick={0,0.25,0.5,0.75,1.0},
            xlabel={\scriptsize $\theta_1$},
            ylabel={\scriptsize $\theta_2$},
            ] 
            \addplot3+[mesh,scatter,only marks,mark size=1.5pt] table[x=theta1, y=theta2, z=g, col sep=comma] {csv/reciprocity_ultimatum/reciprocity_swne_filtered.txt};
        \end{axis}
        \end{tikzpicture}        
    \end{subfigure}
    \hspace{-1.0cm}
    \begin{subfigure}{0.3\textwidth}
        \centering 
        \begin{tikzpicture}
        \begin{axis}[
            title={\scriptsize $u_1$},
            width=\textwidth,
            grid=major,
            view={-135}{20},
            zmin=5,
            zmax=13,
            ztick={0,5,9,13},
            xlabel={\scriptsize $\theta_1$},
            ylabel={\scriptsize $\theta_2$},
            ] 
            \addplot3+[mesh,scatter,only marks,mark size=1.5pt] table[x=theta1, y=theta2, z=p1, col sep=comma] {csv/reciprocity_ultimatum/reciprocity_swne_filtered.txt};
        \end{axis}
        \end{tikzpicture}        
    \end{subfigure}
    \hspace{-1.0cm}
    \begin{subfigure}{0.3\textwidth}
        \centering 
        \begin{tikzpicture}
        \begin{axis}[
            title={\scriptsize $u_2$},
            width=\textwidth,
            grid=major,
            view={-135}{20},
            zmin=5,
            zmax=13,
            ztick={0,5,9,13},
            xlabel={\scriptsize $\theta_1$},
            ylabel={\scriptsize $\theta_2$},
            ] 
            \addplot3+[mesh,scatter,only marks,mark size=1.5pt] table[x=theta1, y=theta2, z=p2, col sep=comma] {csv/reciprocity_ultimatum/reciprocity_swne_filtered.txt};
        \end{axis}
        \end{tikzpicture}        
    \end{subfigure}
    \begin{subfigure}{0.3\textwidth}
        \centering 
        \begin{tikzpicture}
        \begin{axis}[
            title={\scriptsize $\probact{g}$},
            width=\textwidth,
            grid=major,
            view={-135}{20},
            zmin=0.0,
            zmax=1.0,
            ztick={0,0.25,0.5,0.75,1.0},
            xlabel={\scriptsize $\theta_1$},
            ylabel={\scriptsize $\theta_2$},
            ] 
            \addplot3+[mesh,scatter,only marks,mark size=1.5pt] table[x=theta1, y=theta2, z=r, col sep=comma] {csv/reciprocity_ultimatum/ultimatum_swne_filtered.txt};
        \end{axis}
        \end{tikzpicture}        
    \end{subfigure}
    \hspace{-1.0cm}
    \begin{subfigure}{0.3\textwidth}
        \centering 
        \begin{tikzpicture}
        \begin{axis}[
            title={\scriptsize $\probact{a}$},
            width=\textwidth,
            grid=major,
            view={-135}{20},
            zmin=0.0,
            zmax=1.0,
            ztick={0,0.25,0.5,0.75,1.0},
            xlabel={\scriptsize $\theta_1$},
            ylabel={\scriptsize $\theta_2$},
            ] 
            \addplot3+[mesh,scatter,only marks,mark size=1.5pt] table[x=theta1, y=theta2, z=g, col sep=comma] {csv/reciprocity_ultimatum/ultimatum_swne_filtered.txt};
        \end{axis}
        \end{tikzpicture}        
    \end{subfigure}
    \hspace{-1.0cm}    
    \begin{subfigure}{0.3\textwidth}
        \centering 
        \begin{tikzpicture}
        \begin{axis}[
            title={\scriptsize $u_1$},
            width=\textwidth,
            grid=major,
            view={-135}{20},
            zmin=0,
            zmax=14,
            ztick={0,7,14},
            xlabel={\scriptsize $\theta_1$},
            ylabel={\scriptsize $\theta_2$},
            ] 
            \addplot3+[mesh,scatter,only marks,mark size=1.5pt] table[x=theta1, y=theta2, z=p1, col sep=comma] {csv/reciprocity_ultimatum/ultimatum_swne_filtered.txt};
        \end{axis}
        \end{tikzpicture}        
    \end{subfigure}
    \hspace{-1.0cm}
    \begin{subfigure}{0.3\textwidth}
        \centering 
        \begin{tikzpicture}
        \begin{axis}[
            title={\scriptsize $u_2$},
            width=\textwidth,
            grid=major,
            view={-135}{20},
            zmin=0,
            zmax=14,
            ztick={0,7,14},
            xlabel={\scriptsize $\theta_1$},
            ylabel={\scriptsize $\theta_2$},
            ] 
            \addplot3+[mesh,scatter,only marks,mark size=1.5pt] table[x=theta1, y=theta2, z=p2, col sep=comma] {csv/reciprocity_ultimatum/ultimatum_swne_filtered.txt};
        \end{axis}
        \end{tikzpicture}        
    \end{subfigure}
    \caption{Possible social welfare PEs for the reciprocity (top) and ultimatum (bottom) games for a range of reciprocity sensitivities $\theta_1,\theta_2$. We show the strategy (probabilities $\probact{g}$, $\probact{a}$ of choosing $\mathit{greedy}$, $\mathit{accept}$) and values (utilities $u_1$ and $u_2$).}
    \label{fig:reciprocity_ultimatum_swne}
\end{figure}

%% file: figures/crossing/crossing.tex
\tikzset{
    pics/man/.style={code=
    {
    \draw[#1]   
        (0,0) .. controls ++(0,-0.8) and ++(0.2,0.6) ..
        (-0.4,-1.8) .. controls ++(0.2,-0.8) and ++(0.1,0.6) ..
        (-0.5,-4.4) .. controls ++(-0.6,-0.2) and ++(0.7,0.1) ..
        (-2,-4.8) .. controls ++(0,0.3) and ++(-0.5,-0.2)  ..
        (-1,-3.8) .. controls ++(-0.1,0.9) and ++(-0.1,-0.8)  ..
        (-1,-1.8) .. controls ++(-0.3,1) and ++(-0.2,-0.8)  ..              
        (-0.9,0.9) .. controls ++(-0.1,1) and ++(0,-0.8)  ..
        (-1.2,2.8) .. controls ++(-0.4,-1) and ++(0.4,0.5)  ..
        (-2.6,0.8) .. controls ++(0.5,-0.8) and ++(0.2,-0.1)  ..
        (-3.2,-0.1) .. controls ++(-0.2,0) and ++(-0.3,-0.5)  ..
        (-3.3,0.8) .. controls ++(0.4,0.5) and ++(-0.5,-0.5)  ..        
        (-1.8,3.4) .. controls ++(0.5,0.5) and ++(-0.3,-0.1)  ..                
        (-0.7,3.9) .. controls ++(0.3,0.1) and ++(0,-0.2)  ..
        (-0.4,4.3) .. controls ++(-1.2,0.3) and ++(-1.2,0)  ..
        (0,6.2) coordinate (-head) .. controls ++(1.2,0) and ++(1.2,0.3) .. 
        (0.4,4.3) .. controls ++(0,-0.2) and ++(-0.3,0.1) ..
        (0.7,3.9) .. controls ++(0.3,-0.1) and ++(-0.5,0.5) ..
        (1.8,3.4) .. controls ++(0.5,-0.5) and ++(-0.4,0.5) ..
        (3.3,0.8) .. controls ++(0.3,-0.5) and ++(0.2,0) ..
        (3.2,-0.1) .. controls ++(-0.2,-0.1) and ++(-0.5,-0.8) ..
        (2.6,0.8) .. controls ++(-0.4,0.5) and ++(0.4,-1) ..
        (1.2,2.8) .. controls ++(0,-0.8) and ++(0.1,1) ..
        (0.9,0.9) .. controls ++(0.2,-0.8) and ++(0.3,1) ..
        (1,-1.8) .. controls ++(0.1,-0.8) and ++(0.1,0.9) ..
        (1,-3.8) .. controls ++(0.5,-0.2) and ++(0,0.3) ..
        (2,-4.8) .. controls ++(-0.7,0.1) and ++(0.6,-0.2) ..
        (0.5,-4.4) .. controls ++(-0.1,0.6) and ++(-0.2,-0.8) ..
        (0.4,-1.8) .. controls ++(-0.2,0.6) and ++(0,-0.8) ..
        (0,0) ++ (0,2) coordinate (-heart) -- cycle
        ;
    },
    }
}

\newcommand{\customlatpath}{(-0.092,1.65) to[out=0,in=180] (0.105,1.65)}

\begin{figure}[t]
\centering
\hspace{1.00cm}
    \begin{subfigure}{0.48\textwidth}
    \begin{tikzpicture}[scale=4.0]
        \clip (-0.6,0.2) rectangle (0.6,1.8);
        \path[fill=green!50!black] (-1,0) rectangle (1,2.5);
        \draw[line width=25,gray] \custompath;
        \draw[draw=white,dashed,double=white,double distance=20] \custompath;
        \draw[line width=20,gray] \custompath;
        \draw[white,decorate,decoration={street mark},street mark distance=13] \custompath;
        \draw[white,decorate,decoration={street mark},street mark distance=-13] \custompath;
        
        \draw[draw=brown,dashed,double=brown,double distance=6, xshift=5.5] \custompath;
        \draw[brown,decorate,decoration={street mark},street mark distance=-20] \custompath;
        \draw[draw=brown,dashed,double=brown,double distance=6, xshift=-5.5] \custompath;
        \draw[brown,decorate,decoration={street mark},street mark distance=20] \custompath;
    
        \draw[draw=white,dashed,double=white,double distance=18] \customlatpath;
    
        \draw[decorate,decoration={markings,
         mark=at position 0.1 with {\draw[cyan,-latex,line width=2pt] (0.25,0)
         coordinate (0.25,0) -- (1.25,0);}
        }] \custompath;
        \node[sedan top,body color=red!70,window color=black!80,minimum width=0.9cm,rotate = 90,scale=1.2] (car) at
        (0,0.35) {};
    
        \def\xp{0.2}
        \def\yp{1.45}
    
        \draw[magenta,-latex,line width=2pt] (\xp-0.05,\yp) -> (\xp-0.2,\yp);
        
        \draw (\xp,\yp) pic(M){man={scale=0.05,black!50!black,fill=black}};
    
        \node[draw,circle,fill=yellow,inner sep=2pt] at (0,0.75) (p1) {};
        \node[draw,circle,fill=yellow,inner sep=2pt] at (0,1.0) (p2) {};
        \node[draw,circle,fill=orange,inner sep=2pt] at (-0.05,1.25) (p3) {};
        \node[draw,circle,fill=purple,inner sep=2pt] at (0.05,1.25) (p4) {};
    
        \node[] at (-0.4,0.925) (av) {\scriptsize \shortstack[c]{maintain/ \\ reduce}};
    
        \node[] at (\xp+0.2,\yp) (ap) {\scriptsize \shortstack[c]{cross/ \\ wait}};
    
        \draw[yellow,thick] (p1) -- (av.east);
        \draw[yellow,thick] (p2) -- (av.east);
    
        \draw[->,dotted] (p1) -- (p2);
        \draw[->,dotted] (p2) -- (p3) node [pos=0.5,xshift=-0.15cm] {\scriptsize $\gamma$};
        \draw[->,dotted] (p2) -- (p4) node [pos=0.5,xshift=0.3cm] {\scriptsize $1{-}\gamma$};
    \end{tikzpicture}
    \end{subfigure}
    \caption{An illustration for the pedestrian crossing scenario.}
    \label{fig:pedestrian_crossing}
\end{figure}
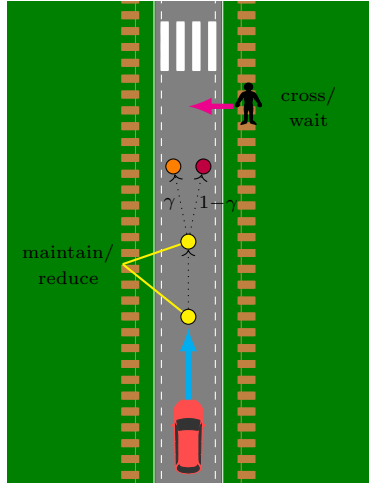

%% file: figures/crossing/crossing_alpha_5.tex
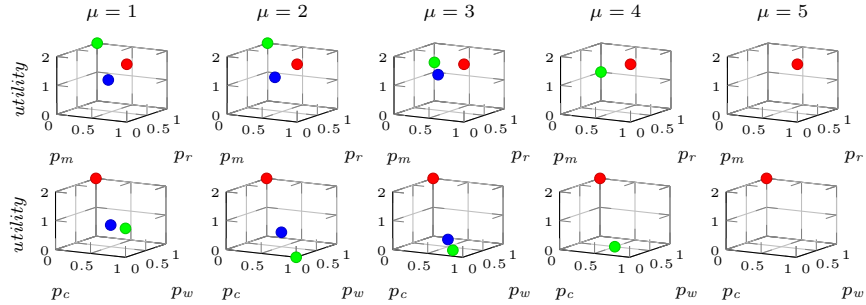
\begin{figure}[t]
\tiny
\centering
\begin{subfigure}{0.25\textwidth}
\centering
    \begin{tikzpicture}
        \begin{axis}[
            colormap={seq}{
                color=(red);
                color=(green);
                color=(blue);
            },
            point meta min = 0,
            point meta max = 2,
            point meta={\coordindex},
            title={\scriptsize $\mu=1$},
            width=\textwidth,
            grid=major,
            view={30}{15},
            zmin=0.0,
            zmax=2.0,
            ztick={0.0,1.0,2.0},
            xtick={0.0,0.5,1.0},
            ytick={0.0,0.5,1.0},
            xmin=0.0,xmax=1.0,
            ymin=0.0,ymax=1.0,
            xlabel={\scriptsize $\probact{m}$},
            ylabel={\scriptsize $\probact{r}$},
            zlabel={\scriptsize \emph{utility}}
            ]
            \addplot3+[mesh,scatter,only marks,mark size=2pt] table[x=m, y=r, z=car,col sep=comma] {csv/crossing/crossing_1.txt};
        \end{axis}
    \end{tikzpicture}    
\end{subfigure}
\hspace{-.75cm}
\begin{subfigure}{0.25\textwidth}
\centering
    \begin{tikzpicture}
        \begin{axis}[
            colormap={seq}{
                color=(red);
                color=(green);
                color=(blue);
            },
            point meta min = 0,
            point meta max = 2,
            point meta={\coordindex},
            title={\scriptsize $\mu=2$},
            width=\textwidth,
            grid=major,
            view={30}{15},
            zmin=0.0,
            zmax=2.0,
            ztick={0.0,1.0,2.0},
            xtick={0.0,0.5,1.0},
            ytick={0.0,0.5,1.0},
            xmin=0.0,xmax=1.0,
            ymin=0.0,ymax=1.0,
            xlabel={\scriptsize $\probact{m}$},
            ylabel={\scriptsize $\probact{r}$},
            ]         
            \addplot3+[mesh,scatter,only marks,mark size=2pt] table[x=m, y=r, z=car,col sep=comma] {csv/crossing/crossing_2.txt};
        \end{axis}
    \end{tikzpicture}    
\end{subfigure}
\hspace{-1.0cm}
\begin{subfigure}{0.25\textwidth}
\centering
    \begin{tikzpicture}
        \begin{axis}[
            colormap={seq}{
                color=(red);
                color=(green);
                color=(blue);
            },
            point meta min = 0,
            point meta max = 2,
            point meta={\coordindex},
            title={\scriptsize $\mu=3$},
            width=\textwidth,
            grid=major,
            view={30}{15},
            zmin=0.0,
            zmax=2.0,
            ztick={0.0,1.0,2.0},
            xtick={0.0,0.5,1.0},
            ytick={0.0,0.5,1.0},
            xmin=0.0,xmax=1.0,
            ymin=0.0,ymax=1.0,
            xlabel={\scriptsize $\probact{m}$},
            ylabel={\scriptsize $\probact{r}$},
            ]
            \addplot3+[mesh,scatter,only marks,mark size=2pt] table[x=m, y=r, z=car,col sep=comma] {csv/crossing/crossing_3.txt};
        \end{axis}
    \end{tikzpicture}    
\end{subfigure}
\hspace{-1.0cm}
\begin{subfigure}{0.25\textwidth}
\centering
    \begin{tikzpicture}
        \begin{axis}[
            colormap={seq}{
                color=(red);
                color=(green);
                color=(blue);
            },
            point meta min = 0,
            point meta max = 2,
            point meta={\coordindex},
            title={\scriptsize $\mu=4$},
            width=\textwidth,
            grid=major,
            view={30}{15},
            zmin=0.0,
            zmax=2.0,
            ztick={0.0,1.0,2.0},
            xtick={0.0,0.5,1.0},
            ytick={0.0,0.5,1.0},
            xmin=0.0,xmax=1.0,
            ymin=0.0,ymax=1.0,
            xlabel={\scriptsize $\probact{m}$},
            ylabel={\scriptsize $\probact{r}$},
            ]
            \addplot3+[mesh,scatter,only marks,mark size=2pt] table[x=m, y=r, z=car,col sep=comma] {csv/crossing/crossing_4.txt};
        \end{axis}
    \end{tikzpicture}    
\end{subfigure}
\hspace{-1.0cm}
\begin{subfigure}{0.25\textwidth}
\centering
    \begin{tikzpicture}
        \begin{axis}[
            colormap={seq}{
                color=(red);
                color=(green);
                color=(blue);
            },
            point meta min = 0,
            point meta max = 2,
            point meta={\coordindex},
            title={\scriptsize $\mu=5$},
            width=\textwidth,
            grid=major,
            view={30}{15},
            zmin=0.0,
            zmax=2.0,
            ztick={0.0,1.0,2.0},
            xtick={0.0,0.5,1.0},
            ytick={0.0,0.5,1.0},
            xmin=0.0,xmax=1.0,
            ymin=0.0,ymax=1.0,
            xlabel={\scriptsize $\probact{m}$},
            ylabel={\scriptsize $\probact{r}$},
            ]
            \addplot3+[mesh,scatter,only marks,mark size=2pt] table[x=m, y=r, z=car,col sep=comma] {csv/crossing/crossing_5.txt};
        \end{axis}
    \end{tikzpicture}    
\end{subfigure}
\begin{subfigure}{0.25\textwidth}
\centering
    \begin{tikzpicture}
        \begin{axis}[
            colormap={seq}{
                color=(red);
                color=(green);
                color=(blue);
            },
            point meta min = 0,
            point meta max = 2,
            point meta={\coordindex},
            width=\textwidth,
            grid=major,
            view={30}{15},
            zmin=0.0,
            zmax=2.0,
            ztick={0.0,1.0,2.0},
            xtick={0.0,0.5,1.0},
            ytick={0.0,0.5,1.0},
            xmin=0.0,xmax=1.0,
            ymin=0.0,ymax=1.0,
            xlabel={\scriptsize $\probact{c}$},
            ylabel={\scriptsize $\probact{w}$},
            zlabel={\scriptsize \emph{utility}}
            ]
            \addplot3+[mesh,scatter,only marks,mark size=2pt] table[x=c, y=w, z=ped,col sep=comma] {csv/crossing/crossing_1.txt};
        \end{axis}
    \end{tikzpicture}    
\end{subfigure}
\hspace{-.75cm}
\begin{subfigure}{0.25\textwidth}
\centering
    \begin{tikzpicture}
        \begin{axis}[
            colormap={seq}{
                color=(red);
                color=(green);
                color=(blue);
            },
            point meta min = 0,
            point meta max = 2,
            point meta={\coordindex},        
            width=\textwidth,
            grid=major,
            view={30}{15},
            zmin=0.0,
            zmax=2.0,
            ztick={0.0,1.0,2.0},
            xtick={0.0,0.5,1.0},
            ytick={0.0,0.5,1.0},
            xmin=0.0,xmax=1.0,
            ymin=0.0,ymax=1.0,
            xlabel={\scriptsize $\probact{c}$},
            ylabel={\scriptsize $\probact{w}$},
            ]         
            \addplot3+[mesh,scatter,only marks,mark size=2pt] table[x=c, y=w, z=ped,col sep=comma] {csv/crossing/crossing_2.txt};
        \end{axis}
    \end{tikzpicture}    
\end{subfigure}
\hspace{-1.0cm}
\begin{subfigure}{0.25\textwidth}
\centering
    \begin{tikzpicture}
        \begin{axis}[
            colormap={seq}{
                color=(red);
                color=(green);
                color=(blue);
            },
            point meta min = 0,
            point meta max = 2,
            point meta={\coordindex},        
            width=\textwidth,
            grid=major,
            view={30}{15},
            zmin=0.0,
            zmax=2.0,
            ztick={0.0,1.0,2.0},
            xtick={0.0,0.5,1.0},
            ytick={0.0,0.5,1.0},
            xmin=0.0,xmax=1.0,
            ymin=0.0,ymax=1.0,
            xlabel={\scriptsize $\probact{c}$},
            ylabel={\scriptsize $\probact{w}$},
            ]
            \addplot3+[mesh,scatter,only marks,mark size=2pt] table[x=c, y=w, z=ped,col sep=comma] {csv/crossing/crossing_3.txt};
        \end{axis}
    \end{tikzpicture}    
\end{subfigure}
\hspace{-1.0cm}
\begin{subfigure}{0.25\textwidth}
\centering
    \begin{tikzpicture}
        \begin{axis}[
            colormap={seq}{
                color=(red);
                color=(green);
                color=(blue);
            },
            point meta min = 0,
            point meta max = 2,
            point meta={\coordindex},  
            width=\textwidth,
            grid=major,
            view={30}{15},
            zmin=0.0,
            zmax=2.0,
            ztick={0.0,1.0,2.0},
            xtick={0.0,0.5,1.0},
            ytick={0.0,0.5,1.0},
            xmin=0.0,xmax=1.0,
            ymin=0.0,ymax=1.0,
            xlabel={\scriptsize $\probact{c}$},
            ylabel={\scriptsize $\probact{w}$},
            ]
            \addplot3+[mesh,scatter,only marks,mark size=2pt] table[x=c, y=w, z=ped,col sep=comma] {csv/crossing/crossing_4.txt};
        \end{axis}
    \end{tikzpicture}    
\end{subfigure}
\hspace{-1.0cm}
\begin{subfigure}{0.25\textwidth}
\centering
    \begin{tikzpicture}
        \begin{axis}[
            colormap={seq}{
                color=(red);
                color=(green);
                color=(blue);
            },
            point meta min = 0,
            point meta max = 2,
            point meta={\coordindex},        
            width=\textwidth,
            grid=major,
            view={30}{15},
            zmin=0.0,
            zmax=2.0,
            ztick={0.0,1.0,2.0},
            xtick={0.0,0.5,1.0},
            ytick={0.0,0.5,1.0},
            xmin=0.0,xmax=1.0,
            ymin=0.0,ymax=1.0,
            xlabel={\scriptsize $\probact{c}$},
            ylabel={\scriptsize $\probact{w}$},
            ]
            \addplot3+[mesh,scatter,only marks,mark size=2pt] table[x=c, y=w, z=ped,col sep=comma] {csv/crossing/crossing_5.txt};
        \end{axis}
    \end{tikzpicture}    
\end{subfigure}
\caption{Equilibria strategies and utilities for the vehicle (top) and pedestrian (bottom) in the pedestrian crossing scenario, for different values of $\mu$.}
\label{fig:eq_cross_alpha}
\end{figure}

%% file: figures/cyclist_vehicle/cyclist_vehicle_model.tex
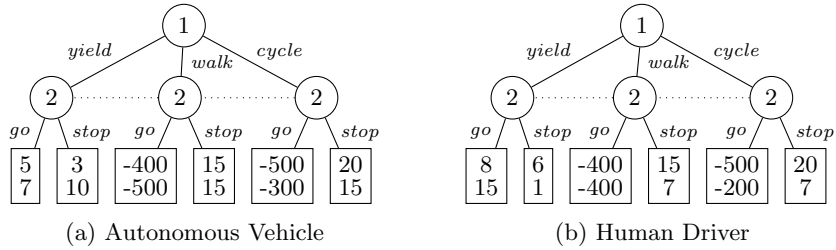
\begin{figure}[t]
\centering
\begin{subfigure}{0.49\textwidth}
\centering
    \begin{forest}
      [$1$, node options={draw,circle}, name=p1
            [$2$, node options={draw,circle}, edge={thin}, edge label={node[pos=0.3,left,xshift=-0.2cm]{\scriptsize \emph{yield}}}, name=p21, 
                [{\shortstack[c]{5 \\ 7}}, node options={draw,rectangle}, name=l1, edge={thin}, edge label={node[midway,left]{\scriptsize \emph{go}}}]
    		[{\shortstack[c]{3 \\ 10}}, node options={draw,rectangle}, name=l2, edge={thin}, edge label={node[midway,right]{\scriptsize \emph{stop}}}]
    	]
            [$2$, node options={draw,circle}, edge={thin}, edge label={node[pos=0.5,right]{\scriptsize \emph{walk}}}, name=p22, tikz={\draw [dotted] () [] to (p21);}
                [{\shortstack[c]{-400 \\ -500}}, node options={draw,rectangle}, name=l3, edge={thin}, edge label={node[midway,left]{\scriptsize \emph{go}}}]
    		[{\shortstack[c]{15 \\ 15}}, node options={draw,rectangle}, name=l4, edge={thin}, edge label={node[midway,right]{\scriptsize \emph{stop}}}]
    	]
            [$2$, node options={draw,circle}, edge={thin}, edge label={node[pos=0.3,right,xshift=0.2cm]{\scriptsize \emph{cycle}}}, name=p23, tikz={\draw [dotted] () [] to (p22);}
                [{\shortstack[c]{-500 \\ -300}}, node options={draw,rectangle}, name=l5, edge={thin}, edge label={node[midway,left]{\scriptsize \emph{go}}}]
    		[{\shortstack[c]{20 \\ 15}}, node options={draw,rectangle}, name=l6, edge={thin}, edge label={node[midway,right]{\scriptsize \emph{stop}}}]
    	]     
      ]
    \end{forest}
    \subcaption{Autonomous Vehicle}
\end{subfigure}
\begin{subfigure}{0.49\textwidth}
\centering
    \begin{forest}
      [$1$, node options={draw,circle}, name=p1
            [$2$, node options={draw,circle}, edge={thin}, edge label={node[pos=0.3,left,xshift=-0.2cm]{\scriptsize \emph{yield}}}, name=p21, 
                [{\shortstack[c]{8 \\ 15}}, node options={draw,rectangle}, name=l1, edge={thin}, edge label={node[midway,left]{\scriptsize \emph{go}}}]
    		[{\shortstack[c]{6 \\ 1}}, node options={draw,rectangle}, name=l2, edge={thin}, edge label={node[midway,right]{\scriptsize \emph{stop}}}]
    	]
            [$2$, node options={draw,circle}, edge={thin}, edge label={node[pos=0.5,right]{\scriptsize \emph{walk}}}, name=p22, tikz={\draw [dotted] () [] to (p21);}
                [{\shortstack[c]{-400 \\ -400}}, node options={draw,rectangle}, name=l3, edge={thin}, edge label={node[midway,left]{\scriptsize \emph{go}}}]
    		[{\shortstack[c]{15 \\ 7}}, node options={draw,rectangle}, name=l4, edge={thin}, edge label={node[midway,right]{\scriptsize \emph{stop}}}]
    	]
            [$2$, node options={draw,circle}, edge={thin}, edge label={node[pos=0.3,right,xshift=0.2cm]{\scriptsize \emph{cycle}}}, name=p23, tikz={\draw [dotted] () [] to (p22);}
                [{\shortstack[c]{-500 \\ -200}}, node options={draw,rectangle}, name=l5, edge={thin}, edge label={node[midway,left]{\scriptsize \emph{go}}}]
    		[{\shortstack[c]{20 \\ 7}}, node options={draw,rectangle}, name=l6, edge={thin}, edge label={node[midway,right]{\scriptsize \emph{stop}}}]
    	]     
      ]
    \end{forest}
    \subcaption{Human Driver}
\end{subfigure}
    \caption{Original cyclist vs. vehicle game. In \cite{MB18}, the left and right games are played with probabilities $p$ and $1-p$, respectively. The utilities for the cyclist (1) and the vehicle (2) are given in the top and bottom rows of each leaf node.}
    \label{fig:cyclist_vehicle}
\end{figure}

%% file: figures/cyclist_vehicle/cyclist_vehicle_plot.tex
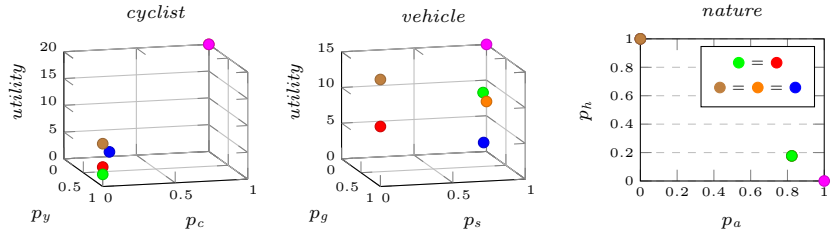
\begin{figure}[t]
\tiny
\centering
\begin{subfigure}{0.33\textwidth}
\centering
    \begin{tikzpicture}
        \begin{axis}[
            colormap={seq}{
                color=(red);
                color=(green);
                color=(blue);
                color=(orange);
                color=(magenta);
                color=(brown);
            },
            title={\scriptsize \emph{cyclist}},
            point meta min=0,
            point meta max=5,
            point meta={\coordindex},
            width=\textwidth,
            grid=major,
            view={75}{15},
            zmin=0.0,
            zmax=20,
            ztick={0,5,10,15,20},
            xtick={0.0,0.5,1.0},
            ytick={0.0,0.5,1.0},
            xmin=0.0,xmax=1.0,
            ymin=0.0,ymax=1.0,
            xlabel={\scriptsize $\probact{y}$},
            ylabel={\scriptsize $\probact{c}$},
            zlabel={\scriptsize \emph{utility}}
            ]
            \addplot3+[mesh,scatter,only marks,mark size=2pt] table[x=y, y=c, z=p1,col sep=comma] {csv/cyclist_vehicle/cyclist_vehicle.txt};
        \end{axis}
    \end{tikzpicture}    
\end{subfigure}
\hspace{-0.5cm}
\begin{subfigure}{0.33\textwidth}
\centering
    \begin{tikzpicture}
        \begin{axis}[
            colormap={seq}{
                color=(red);
                color=(green);
                color=(blue);
                color=(orange);
                color=(magenta);
                color=(brown);
            },
            title={\scriptsize \emph{vehicle}},
            point meta min=0,
            point meta max=5,
            point meta={\coordindex},
            width=\textwidth,
            grid=major,
            view={75}{15},
            zmin=0.0,
            zmax=15,
            ztick={0,5,10,15},
            xtick={0.0,0.5,1.0},
            ytick={0.0,0.5,1.0},
            xmin=0.0,xmax=1.0,
            ymin=0.0,ymax=1.0,
            xlabel={\scriptsize $\probact{g}$},
            ylabel={\scriptsize $\probact{s}$},
            zlabel={\scriptsize \emph{utility}}
            ]
            \addplot3+[mesh,scatter,only marks,mark size=2pt] table[x=g, y=s, z=p2,col sep=comma] {csv/cyclist_vehicle/cyclist_vehicle.txt};
        \end{axis}
    \end{tikzpicture}    
\end{subfigure}
\hspace{-0.25cm}
\begin{subfigure}{.33\textwidth}
    \centering
    \tiny {
    \begin{tikzpicture}
    \begin{axis}[
        colormap={seq}{
                color=(red);
                color=(green);
                color=(blue);
                color=(orange);
                color=(magenta);
                color=(brown);
        },
        title={\scriptsize \emph{nature}},
        point meta min=0,
        point meta max=5,
        point meta={\coordindex},
        ylabel={\scriptsize $\probact{h}$},
        y label style={at={(axis description                                    cs:0.275,0.5)},anchor=south
        },
        xlabel={\scriptsize $\probact{a}$},
        x label style={at={(axis description                                    cs:0.5,-0.1)},anchor=south
        },
        xmin=0.0, xmax=1.0,
        xtick={0.0,0.2,0.4,0.6,0.8,1.0},
        ymin=0.0, ymax=1.0,
        ytick={0.0,0.2,0.4,0.6,0.8,1.0},
        ymajorgrids=true,
        grid style=dashed,
        width=\textwidth,
        legend entries={
                {\tikz\draw[draw,green,fill=green] (0,0) circle (2pt); \raisebox{0.035cm}{$=$} \tikz\draw[draw,red,fill=red] (0,0) circle (2pt);},
                {\tikz\draw[draw,brown,fill=brown] (0,0) circle (2pt); \raisebox{0.035cm}{$=$} \tikz\draw[draw,orange,fill=orange] (0,0) circle (2pt); \raisebox{0.035cm}{$=$} \tikz\draw[draw,blue,fill=blue] (0,0) circle (2pt);}},
        legend style={at={(0.95,0.95)},
                    anchor=north east}            
    ]
    \addlegendimage{empty legend}
    \addlegendimage{empty legend}
    ]
    \addplot[mark=*,mark size=2.0pt,only marks,scatter] table [x=a, y=h, col sep=comma]{csv/cyclist_vehicle/cyclist_vehicle.txt};
    \end{axis}
    \end{tikzpicture}
    }
    \end{subfigure}
    \caption{Strategies and utilities for the cyclist vs. vehicle game.}
    \label{fig:eq_cyclist_vehicle_plot}
\end{figure}

%% file: figures/multistage_crossing/multistage_crossing_model.tex
\begin{figure}[t]
\centering
\begin{tikzpicture}[->,>=latex,auto,node distance=2.8cm, semithick, scale=.40]
 \tikzstyle{every state}=[draw=black, text=black, initial text={}]

\small
\node[initial above, state, inner sep=2pt, minimum size=0pt, label={right:{\scriptsize $\langle 0,0,0 \rangle$}}] (S)at(0,2.75) (s0) {\scriptsize $s_0$}; 
\node[state, inner sep=2pt, minimum size=0pt, label={left:{\scriptsize $\langle 1,0,0 \rangle$}}] (S)at(-7,-3) (s1) {\scriptsize $s_1$};
\node[state, inner sep=2pt, minimum size=0pt, label={left:{\scriptsize $\langle 1,0,1 \rangle$}}] (S)at(-2,-3) (s2) {\scriptsize $s_2$}; 
\node[state, inner sep=2pt, minimum size=0pt, label={right:{\scriptsize $\langle 1,1,0 \rangle$}}] (S)at(2,-3) (s3) {\scriptsize $s_3$}; 
\node[state, inner sep=2pt, minimum size=0pt, label={right:{\scriptsize $\langle 1,1,1 \rangle$}}] (S)at(7,-3) (s4) {\scriptsize $s_4$}; 

\node[draw, circle, fill, black, inner sep=1pt] at(0,1.25) (d) {};

\path [-] (s0.south) [] edge node[left, xshift=-1mm] {\scriptsize $(w,r)$} (d);

\path [->] (d) [bend right] edge node[left, xshift=-1mm, pos=0.8] {\scriptsize $(1{-}\gamma){\cdot}(1{-}\gamma)$} (s1.north);

\path [->] (d) [] edge node[left, xshift=-1mm, pos=0.5] {\scriptsize $\gamma{\cdot}(1{-}\gamma)$} (s2.north);

\path [->] (d) [] edge node[right, xshift=0mm, pos=0.5] {\scriptsize $\gamma{\cdot}(1{-}\gamma)$} (s3.north);

\path [->] (d) [bend left] edge node[right, xshift=1mm, pos=0.8] {\scriptsize $\gamma{\cdot}\gamma$} (s4.north);

\end{tikzpicture}
\caption{Transitions for actions $(w,r)$ from the initial state of the CSG for the multi-stage pedestrian crossing example. States are of the form $\langle j,c_r,c_w \rangle$, for step count $j$ and variables $c_r$, $c_w$.}
\label{fig:pedestrian_transitions}
\end{figure}
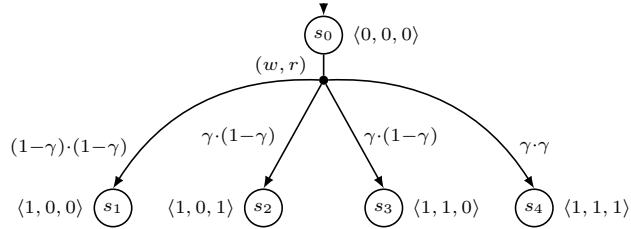

%% file: figures/multistage_crossing/multistage_crossing_plot.tex
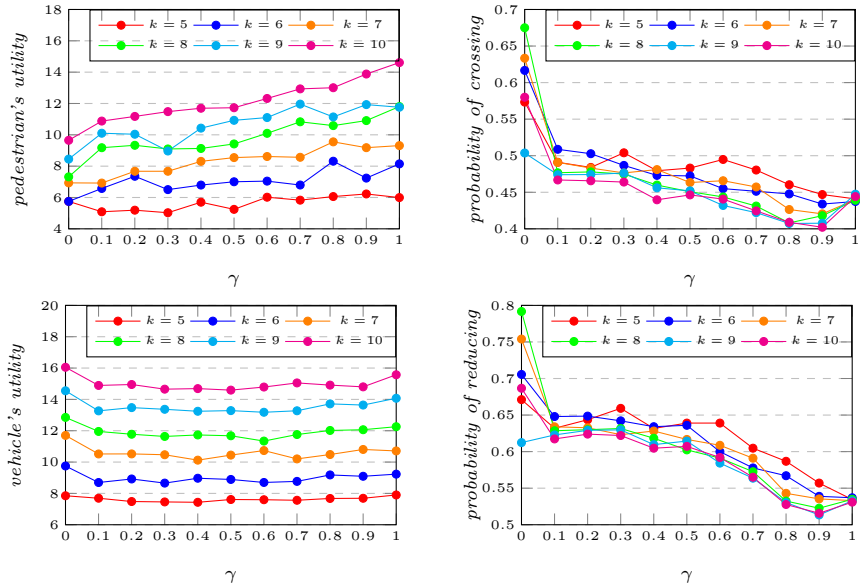
\begin{figure}[t]
    \tiny
    \centering
    \begin{subfigure}{0.49\textwidth}
        \centering
        \begin{tikzpicture}
        \begin{axis}[
            ylabel={\scriptsize \emph{utilities}},
            xlabel={\scriptsize $\gamma$},
            xmin=0, xmax=1,
            xtick={0.0,0.1,0.2,0.3,0.4,0.5,0.6,0.7,0.8,0.9,1.0},
            ylabel={\scriptsize \emph{pedestrian's utility}},
            y label style={at={(axis description                                    cs:0.175,0.5)},anchor=south
            },
            ymin=4, ymax=18,
            ytick={4,6,8,10,12,14,16,18},
            legend columns=3,
            legend pos=north east,
            legend style={fill=none, 
                        at={(1,1)}, anchor=north east, 
                        nodes={scale=0.9, transform shape},
                        cells={align=left}},
            ymajorgrids=true,
            grid style=dashed,
            height=4.5cm,
            width=\textwidth,
            legend entries={
                    $k=5$,
                    $k=6$,
                    $k=7$,
                    $k=8$,      
                    $k=9$,      
                    $k=10$,      
                    }]
        \addlegendimage{mark=*,red,mark size=1.5pt}
        \addlegendimage{mark=*,blue,mark size=1.5pt}
        \addlegendimage{mark=*,orange,mark size=1.5pt}
        \addlegendimage{mark=*,green,mark size=1.5pt}
        \addlegendimage{mark=*,cyan,mark size=1.5pt}
        \addlegendimage{mark=*,magenta,mark size=1.5pt}
        ]
        \addplot[mark=*,red,mark size=1.5pt] table [x=beta, y=p2, col sep=comma]{csv/multistage_crossing/crossing_multi_filtered_e_5.txt};
        \addplot[mark=*,blue,mark size=1.5pt] table [x=beta, y=p2, col sep=comma]{csv/multistage_crossing/crossing_multi_filtered_e_6.txt};
        \addplot[mark=*,orange,mark size=1.5pt] table [x=beta, y=p2, col sep=comma]{csv/multistage_crossing/crossing_multi_filtered_e_7.txt};
        \addplot[mark=*,green,mark size=1.5pt] table [x=beta, y=p2, col sep=comma]{csv/multistage_crossing/crossing_multi_filtered_e_8.txt};
        \addplot[mark=*,cyan,mark size=1.5pt] table [x=beta, y=p2, col sep=comma]{csv/multistage_crossing/crossing_multi_filtered_e_9.txt};
        \addplot[mark=*,magenta,mark size=1.5pt] table [x=beta, y=p2, col sep=comma]{csv/multistage_crossing/crossing_multi_filtered_e_10.txt};
        \end{axis}
        \end{tikzpicture}
    \end{subfigure}
    \begin{subfigure}{0.49\textwidth}
    \centering
        \begin{tikzpicture}
        \begin{axis}[
            ylabel={\scriptsize \emph{utilities}},
            xlabel={\scriptsize $\gamma$},
            ymin=0.4, ymax=0.7,
            ytick={0.4,0.45,0.5,0.55,0.6,0.65,0.7},
            ylabel={\scriptsize \emph{probability of crossing}},
            y label style={at={(axis description                                    cs:0.175,0.5)},anchor=south
            },
            xmin=0, xmax=1,
            xtick={0.0,0.1,0.2,0.3,0.4,0.5,0.6,0.7,0.8,0.9,1.0},
            legend columns=3,
            legend pos=north east,
            legend style={fill=none, 
                        at={(1,1)}, anchor=north east, 
                        nodes={scale=0.9, transform shape},
                        cells={align=left}},
            ymajorgrids=true,
            grid style=dashed,
            height=4.5cm,
            width=\textwidth,
            legend entries={
                    $k=5$,
                    $k=6$,
                    $k=7$,
                    $k=8$,      
                    $k=9$,      
                    $k=10$,      
                    }]
        \addlegendimage{mark=*,red,mark size=1.5pt}
        \addlegendimage{mark=*,blue,mark size=1.5pt}
        \addlegendimage{mark=*,orange,mark size=1.5pt}
        \addlegendimage{mark=*,green,mark size=1.5pt}
        \addlegendimage{mark=*,cyan,mark size=1.5pt}
        \addlegendimage{mark=*,magenta,mark size=1.5pt}
        ]
        \addplot[mark=*,red,mark size=1.5pt] table [x=beta, y=c, col sep=comma]{csv/multistage_crossing/crossing_multi_filtered_s_5.txt};
        \addplot[mark=*,blue,mark size=1.5pt] table [x=beta, y=c, col sep=comma]{csv/multistage_crossing/crossing_multi_filtered_s_6.txt};
        \addplot[mark=*,orange,mark size=1.5pt] table [x=beta, y=c, col sep=comma]{csv/multistage_crossing/crossing_multi_filtered_s_7.txt};
        \addplot[mark=*,green,mark size=1.5pt] table [x=beta, y=c, col sep=comma]{csv/multistage_crossing/crossing_multi_filtered_s_8.txt};
        \addplot[mark=*,cyan,mark size=1.5pt] table [x=beta, y=c, col sep=comma]{csv/multistage_crossing/crossing_multi_filtered_s_9.txt};
        \addplot[mark=*,magenta,mark size=1.5pt] table [x=beta, y=c, col sep=comma]{csv/multistage_crossing/crossing_multi_filtered_s_10.txt};
        \end{axis}
        \end{tikzpicture}        
    \end{subfigure}

    \begin{subfigure}{0.49\textwidth}
    \centering
    \begin{tikzpicture}
    \begin{axis}[
        ylabel={\scriptsize \emph{utilities}},
        xlabel={\scriptsize $\gamma$},
        xmin=0, xmax=1,
        xtick={0.0,0.1,0.2,0.3,0.4,0.5,0.6,0.7,0.8,0.9,1.0},
        ylabel={\scriptsize \emph{vehicle's utility}},
        y label style={at={(axis description                                    cs:0.175,0.5)},anchor=south
        },
        ymin=6, ymax=20,
        ytick={2,4,6,8,10,12,14,16,18,20},
        legend columns=3,
        legend pos=north west,
        legend style={fill=none, 
                    at={(1,1)}, anchor=north east, 
                    nodes={scale=0.9, transform shape},
                    cells={align=left}},
        ymajorgrids=true,
        grid style=dashed,
        height=4.5cm,
        width=\textwidth,
        legend entries={
                $k=5$,
                $k=6$,
                $k=7$,
                $k=8$,      
                $k=9$,      
                $k=10$,      
                }]
    \addlegendimage{mark=*,red,mark size=1.5pt}
    \addlegendimage{mark=*,blue,mark size=1.5pt}
    \addlegendimage{mark=*,orange,mark size=1.5pt}
    \addlegendimage{mark=*,green,mark size=1.5pt}
    \addlegendimage{mark=*,cyan,mark size=1.5pt}
    \addlegendimage{mark=*,magenta,mark size=1.5pt}
    ]
    \addplot[mark=*,red,mark size=1.5pt] table [x=beta, y=p1, col sep=comma]{csv/multistage_crossing/crossing_multi_filtered_e_5.txt};
    \addplot[mark=*,blue,mark size=1.5pt] table [x=beta, y=p1, col sep=comma]{csv/multistage_crossing/crossing_multi_filtered_e_6.txt};
    \addplot[mark=*,orange,mark size=1.5pt] table [x=beta, y=p1, col sep=comma]{csv/multistage_crossing/crossing_multi_filtered_e_7.txt};
    \addplot[mark=*,green,mark size=1.5pt] table [x=beta, y=p1, col sep=comma]{csv/multistage_crossing/crossing_multi_filtered_e_8.txt};
    \addplot[mark=*,cyan,mark size=1.5pt] table [x=beta, y=p1, col sep=comma]{csv/multistage_crossing/crossing_multi_filtered_e_9.txt};
    \addplot[mark=*,magenta,mark size=1.5pt] table [x=beta, y=p1, col sep=comma]{csv/multistage_crossing/crossing_multi_filtered_e_10.txt};
    \end{axis}
    \end{tikzpicture}
    \end{subfigure}
    \begin{subfigure}{0.49\textwidth}
    \centering
    \begin{tikzpicture}
    \begin{axis}[
        ylabel={\scriptsize \emph{utilities}},
        xlabel={\scriptsize $\gamma$},
        xmin=0, xmax=1,
        ytick={0.5,0.55,0.6,0.65,0.7,0.75,0.8},
        ylabel={\scriptsize \emph{probability of reducing}},
        y label style={at={(axis description                                    cs:0.175,0.5)},anchor=south
        },
        ymin=0.5, ymax=0.8,
        xtick={0.0,0.1,0.2,0.3,0.4,0.5,0.6,0.7,0.8,0.9,1.0},
        legend columns=3,
        legend pos=north east,
        legend style={fill=none, 
                    at={(1,1)}, anchor=north east, 
                    nodes={scale=0.9, transform shape},
                    cells={align=left}},
        ymajorgrids=true,
        grid style=dashed,
        height=4.5cm,
        width=\textwidth,
        legend entries={
                $k=5$,
                $k=6$,
                $k=7$,
                $k=8$,      
                $k=9$,      
                $k=10$,      
                }]
    \addlegendimage{mark=*,red,mark size=1.5pt}
    \addlegendimage{mark=*,blue,mark size=1.5pt}
    \addlegendimage{mark=*,orange,mark size=1.5pt}
    \addlegendimage{mark=*,green,mark size=1.5pt}
    \addlegendimage{mark=*,cyan,mark size=1.5pt}
    \addlegendimage{mark=*,magenta,mark size=1.5pt}
    ]
    \addplot[mark=*,red,mark size=1.5pt] table [x=beta, y=r, col sep=comma]{csv/multistage_crossing/crossing_multi_filtered_s_5.txt};
    \addplot[mark=*,blue,mark size=1.5pt] table [x=beta, y=r, col sep=comma]{csv/multistage_crossing/crossing_multi_filtered_s_6.txt};
    \addplot[mark=*,orange,mark size=1.5pt] table [x=beta, y=r, col sep=comma]{csv/multistage_crossing/crossing_multi_filtered_s_7.txt};
    \addplot[mark=*,green,mark size=1.5pt] table [x=beta, y=r, col sep=comma]{csv/multistage_crossing/crossing_multi_filtered_s_8.txt};
    \addplot[mark=*,cyan,mark size=1.5pt] table [x=beta, y=r, col sep=comma]{csv/multistage_crossing/crossing_multi_filtered_s_9.txt};
    \addplot[mark=*,magenta,mark size=1.5pt] table [x=beta, y=r, col sep=comma]{csv/multistage_crossing/crossing_multi_filtered_s_10.txt};
    \end{axis}
    \end{tikzpicture}
    \end{subfigure}
    \vspace*{-0.2cm}
    \caption{Utility (left) and  probability averages (right) over sampled equilibria for different path lengths in the multi-stage pedestrian crossing example.}
    \label{fig:crossing_multi}
\end{figure}

%% file: table_multistage_crossing_stats.tex
\begin{table}[t]
\setlength{\tabcolsep}{3.5pt}  
\centering
{
\begin{tabular}{|c||c||c|c|c||c||c|c|c|} \hline
$\gamma \in$ & \multicolumn{1}{c||}{$k$} & States & Transitions & Time(s)
& \multicolumn{1}{c||}{$k$} & States & Transitions & Time(s) \\  \hline
\mbox{$[0,0]$} & \multirow{3}{*}{5} & 6 & 21 & 0.76 & \multirow{3}{*}{8} & 9 & 33 & 1.33 \\ \cline{1-1} \cline{3-5} \cline{7-9}
\mbox{$(0,1)$} & & 91 & 701 & 5.97 & & 285 & 2,806 & 25.7 \\ \cline{1-1} \cline{3-5} \cline{7-9}
\mbox{$[1,1]$} & & 91 & 256 & 4.82 & & 285 & 897 & 23.1 \\ \hline \hline 
\mbox{$[0,0]$} & \multirow{3}{*}{6} & 7 & 25 & 0.83 & \multirow{3}{*}{9} & 10 & 37 & 1.73 \\ \cline{1-1} \cline{3-5} \cline{7-9}
\mbox{$(0,1)$} & & 140 & 1,198 & 9.12 & & 385 & 3,981 & 39.6 \\ \cline{1-1} \cline{3-5} \cline{7-9}
\mbox{$[1,1]$} & & 140 & 413 & 8.45 & & 385 & 1,240 & 35.1 \\ \hline \hline 
\mbox{$[0,0]$} & \multirow{3}{*}{7} & 8 & 29 & 1.06 & \multirow{3}{*}{10} & 11 & 41 & 2.09 \\ \cline{1-1} \cline{3-5} \cline{7-9}
\mbox{$(0,1)$} & & 204 & 1,889 & 15.9 & & 506 & 5,446 & 58.7 \\ \cline{1-1} \cline{3-5} \cline{7-9}
\mbox{$[1,1]$} & & 204 & 624 & 14.5 & & 506 & 1,661 & 51.9 \\ \hline 
\end{tabular}}
\vspace*{0.2cm} 
\caption{Statistics for the (CSG) multi-stage pedestrian crossing case study.}
\label{tab:stats}
\vspace*{-0.3cm}
\end{table}

%% file: conclusions.tex
\section{Conclusions}

We have presented techniques that expand the scope of modelling and verification for game-theoretic probabilistic models. Starting with psychological normal form games,  we proposed an NLP encoding that allows us to compute optimal equilibria for individual supports and, through support enumeration, an overall optimal equilibrium for a given NFPG. We then considered CSGs whose states can be expressed as NFPGs, and developed an algorithm to compute equilibria for such CSGs under some restrictions on the type of the players' beliefs. Finally, we reported on a prototype implementation and showcased novel automated analysis, made possible through our method, for a range of case studies.

Verification of psychological games is still a largely unexplored topic and there is ample room for expansion in theory, practice and applications to problems in computer science. Equilibria computation is a hard problem in general, and algorithms for psychological equilibria suffer from some of the same computational drawbacks as those for Nash or correlated equilibria, in addition to presenting new challenges of their own. The main current limitation is having to rely on enumeration for computing an optimal solution, which could be somewhat mitigated by parallelisation and filtering supports as a precomputation step. Future work includes investigating dynamic psychological games \cite{BD09}, which have the advantage of allowing belief updates but pose new modelling and computational challenges, and considering aspects of coordination and robustness via correlated \cite{Per24} and trembling-hand \cite{Kol92} variants. 

\startpara{Acknowledgements}
This project was funded by the ERC under the European
Union’s Horizon 2020 research and innovation programme
(FUN2MODEL, grant agreement No.834115).

%% file: appendix.tex
\clearpage
\appendix

\section{Appendix}\label{appx}

Below, we include larger, more detailed figures for the \emph{cyclist vs.\ vehicle} and \emph{pedestrian crossing} case studies presented in \sectref{sect:traffic}.

\vspace*{-0.9em}
\input{figures/cyclist_vehicle/cyclist_vehilce_appendix}

\input{figures/crossing/crossing_alpha_5_appendix}

%% file: figures/cyclist_vehicle/cyclist_vehilce_appendix.tex
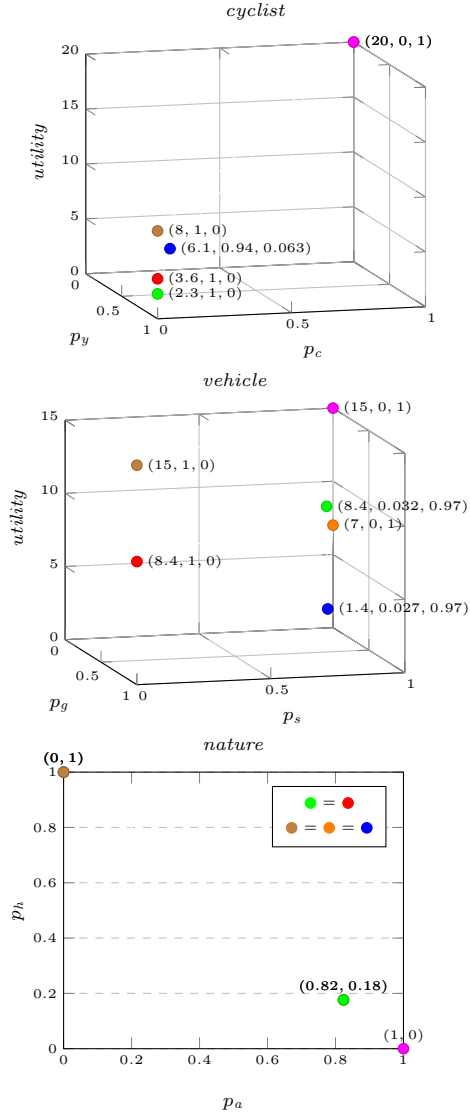
\begin{figure}[!h]
\tiny
\centering
\begin{subfigure}{.5\textwidth}
\centering
    \begin{tikzpicture}
        \begin{axis}[
            colormap={seq}{
                color=(red);
                color=(green);
                color=(blue);
                color=(orange);
                color=(magenta);
                color=(brown);
            },
            title={\scriptsize \emph{cyclist}},
            point meta min=0,
            point meta max=5,
            point meta={\coordindex},
            width=\textwidth,
            grid=major,
            view={75}{12},
            zmin=0.0,
            zmax=20,
            ztick={0,5,10,15,20},
            xtick={0.0,0.5,1.0},
            ytick={0.0,0.5,1.0},
            xmin=0.0,xmax=1.0,
            ymin=0.0,ymax=1.0,
            xlabel={\scriptsize $\probact{y}$},
            ylabel={\scriptsize $\probact{c}$},
            zlabel={\scriptsize \emph{utility}},
            nodes near coords align={horizontal},
            every node near coord/.append style={xshift=0.05cm},
            ]
            \addplot3+[mesh,scatter,only marks,mark size=2pt] table[x=y, y=c, z=p1,col sep=comma] {csv/cyclist_vehicle/cyclist_vehicle.txt};
            \addplot3+[no marks,color=white,opacity=1.0, 
            nodes near coords={\tiny {\color{black}
                $(\pgfmathprintnumber
                    {\pgfkeysvalueof{/data point/z}},
                    \pgfmathprintnumber
                    {\pgfkeysvalueof{/data point/x}},
                    \pgfmathprintnumber
                    {\pgfkeysvalueof{/data point/y}})$%
                }
            }
            ] table[x=y, y=c, z=p1,col sep=comma] {csv/cyclist_vehicle/cyclist_vehicle.txt};
        \end{axis}
    \end{tikzpicture}    
\end{subfigure}

\begin{subfigure}{.5\textwidth}
\centering
    \begin{tikzpicture}
        \begin{axis}[
            colormap={seq}{
                color=(red);
                color=(green);
                color=(blue);
                color=(orange);
                color=(magenta);
                color=(brown);
            },
            title={\scriptsize \emph{vehicle}},
            point meta min=0,
            point meta max=5,
            point meta={\coordindex},
            width=\textwidth,
            grid=major,
            view={75}{12},
            zmin=0.0,
            zmax=15,
            ztick={0,5,10,15},
            xtick={0.0,0.5,1.0},
            ytick={0.0,0.5,1.0},
            xmin=0.0,xmax=1.0,
            ymin=0.0,ymax=1.0,
            xlabel={\scriptsize $\probact{g}$},
            ylabel={\scriptsize $\probact{s}$},
            zlabel={\scriptsize \emph{utility}},
            nodes near coords align={horizontal},
            every node near coord/.append style={xshift=0.05cm}
            ]
            \addplot3+[mesh,scatter,only marks,mark size=2pt] table[x=g, y=s, z=p2,col sep=comma] {csv/cyclist_vehicle/cyclist_vehicle.txt};
            \addplot3+[no marks,color=white,opacity=1.0,
            nodes near coords={\tiny {\color{black}
                $(\pgfmathprintnumber
                    {\pgfkeysvalueof{/data point/z}},
                    \pgfmathprintnumber
                    {\pgfkeysvalueof{/data point/x}},
                    \pgfmathprintnumber
                    {\pgfkeysvalueof{/data point/y}})$%
                }
            }
            ] table[x=g, y=s, z=p2,col sep=comma] {csv/cyclist_vehicle/cyclist_vehicle.txt};
        \end{axis}
    \end{tikzpicture}    
\end{subfigure}

\hspace{-0.4cm}
\begin{subfigure}{.5\textwidth}
    \centering
    \tiny {
    \begin{tikzpicture}
    \begin{axis}[
        colormap={seq}{
                color=(red);
                color=(green);
                color=(blue);
                color=(orange);
                color=(magenta);
                color=(brown);
        },
        title={\scriptsize \emph{nature}},
        point meta min=0,
        point meta max=5,
        point meta={\coordindex},
        ylabel={\scriptsize $\probact{h}$},
        y label style={at={(axis description                                    cs:0.175,0.5)},anchor=south
        },
        xlabel={\scriptsize $\probact{a}$},
        x label style={at={(axis description                                    cs:0.5,-0.1)},anchor=south
        },
        xmin=0.0, xmax=1.0,
        xtick={0.0,0.2,0.4,0.6,0.8,1.0},
        ymin=0.0, ymax=1.0,
        ytick={0.0,0.2,0.4,0.6,0.8,1.0},
        ymajorgrids=true,
        grid style=dashed,
        width=\textwidth,
        legend entries={
                {\tikz\draw[draw,green,fill=green] (0,0) circle (2pt); \raisebox{0.035cm}{$=$} \tikz\draw[draw,red,fill=red] (0,0) circle (2pt);},
                {\tikz\draw[draw,brown,fill=brown] (0,0) circle (2pt); \raisebox{0.035cm}{$=$} \tikz\draw[draw,orange,fill=orange] (0,0) circle (2pt); \raisebox{0.035cm}{$=$} \tikz\draw[draw,blue,fill=blue] (0,0) circle (2pt);}},
        legend style={at={(0.95,0.95)},
                    anchor=north east}            
    ]
    \addlegendimage{empty legend}
    \addlegendimage{empty legend}
    ]
    \addplot[mark=*,mark size=2.0pt,only marks,scatter] table [x=a, y=h, col sep=comma]{csv/cyclist_vehicle/cyclist_vehicle.txt};
    \addplot[no marks,color=white,opacity=1.0,
            nodes near coords={\tiny {\color{black}
                $(\pgfmathprintnumber
                    {\pgfkeysvalueof{/data point/x}},
                    \pgfmathprintnumber
                    {\pgfkeysvalueof{/data point/y}})$%
                }
            }
            ] table[x=a, y=h, col sep=comma] {csv/cyclist_vehicle/cyclist_vehicle.txt};
    \end{axis}
    \end{tikzpicture}
    }
    \end{subfigure}
    \caption{Equilibria strategies and utilities for the cyclist vs.\ vehicle game. Coordinates for the \emph{cyclist}, \emph{vehicle} and \emph{nature} players correspond to $(u,\probact{y},\probact{c})$, $(u,\probact{g},\probact{s})$ and $(\probact{a},\probact{h})$, respectively (where $u$ is the utility). For the cyclist, the utilities and strategies are the same in the orange and magenta equilibria.}
    \label{fig:eq_cyclist_vehicle_plot_appendix}
\end{figure}

%% file: figures/crossing/crossing_alpha_5_appendix.tex
\begin{figure}[h!]
\tiny
\centering
\begin{subfigure}{0.45\textwidth}
\centering
    \begin{tikzpicture}
        \begin{axis}[
            colormap={seq}{
                color=(red);
                color=(green);
                color=(blue);
            },
            point meta min = 0,
            point meta max = 2,
            point meta={\coordindex},
            title={\small $\mu=1$},
            width=\textwidth,
            grid=major,
            view={30}{20},
            zmin=0.0,
            zmax=2.0,
            ztick={0.0,1.0,2.0},
            xtick={0.0,0.5,1.0},
            ytick={0.0,0.5,1.0},
            xmin=0.0,xmax=1.0,
            ymin=0.0,ymax=1.0,
            xlabel={\scriptsize $\probact{m}$},
            ylabel={\scriptsize $\probact{r}$},
            zlabel={\scriptsize \emph{utility}},
            ]
            \addplot3+[mesh,scatter,only marks,mark size=2pt] table[x=m, y=r, z=car,col sep=comma] {csv/crossing/crossing_1.txt};
            \addplot3+[no marks,color=white,opacity=1.0,
            nodes near coords={\tiny {\color{black}
                $(\pgfmathprintnumber
                    {\pgfkeysvalueof{/data point/z}},
                    \pgfmathprintnumber
                    {\pgfkeysvalueof{/data point/x}},
                    \pgfmathprintnumber
                    {\pgfkeysvalueof{/data point/y}})$%
                }
            }
            ] table[x=m, y=r, z=car,col sep=comma] {csv/crossing/crossing_1.txt};
        \end{axis}
    \end{tikzpicture}    
\end{subfigure}
\hfill
\begin{subfigure}{0.45\textwidth}
\centering
    \begin{tikzpicture}
        \begin{axis}[
            colormap={seq}{
                color=(red);
                color=(green);
                color=(blue);
            },
            point meta min = 0,
            point meta max = 2,
            point meta={\coordindex},
            title={\small $\mu=2$},
            width=\textwidth,
            grid=major,
            view={30}{20},
            zmin=0.0,
            zmax=2.0,
            ztick={0.0,1.0,2.0},
            xtick={0.0,0.5,1.0},
            ytick={0.0,0.5,1.0},
            xmin=0.0,xmax=1.0,
            ymin=0.0,ymax=1.0,
            xlabel={\scriptsize $\probact{m}$},
            ylabel={\scriptsize $\probact{r}$},
            ]         
            \addplot3+[mesh,scatter,only marks,mark size=2pt] table[x=m, y=r, z=car,col sep=comma] {csv/crossing/crossing_2.txt};
            \addplot3+[no marks,color=white,opacity=1.0,
            nodes near coords={\tiny {\color{black}
                $(\pgfmathprintnumber
                    {\pgfkeysvalueof{/data point/z}},
                    \pgfmathprintnumber
                    {\pgfkeysvalueof{/data point/x}},
                    \pgfmathprintnumber
                    {\pgfkeysvalueof{/data point/y}})$%
                }
            }
            ] table[x=m, y=r, z=car,col sep=comma] {csv/crossing/crossing_2.txt};
        \end{axis}
    \end{tikzpicture}    
\end{subfigure}


\begin{subfigure}{0.45\textwidth}
\centering
    \begin{tikzpicture}
        \begin{axis}[
            colormap={seq}{
                color=(red);
                color=(green);
                color=(blue);
            },
            point meta min = 0,
            point meta max = 2,
            point meta={\coordindex},
            width=\textwidth,
            grid=major,
            view={30}{20},
            zmin=0.0,
            zmax=2.0,
            ztick={0.0,1.0,2.0},
            xtick={0.0,0.5,1.0},
            ytick={0.0,0.5,1.0},
            xmin=0.0,xmax=1.0,
            ymin=0.0,ymax=1.0,
            xlabel={\scriptsize $\probact{c}$},
            ylabel={\scriptsize $\probact{w}$},
            zlabel={\scriptsize \emph{utility}}
            ]
            \addplot3+[mesh,scatter,only marks,mark size=2pt] table[x=c, y=w, z=ped,col sep=comma] {csv/crossing/crossing_1.txt};
            \addplot3+[no marks,color=white,opacity=1.0,
            nodes near coords={\tiny {\color{black}
                $(\pgfmathprintnumber
                    {\pgfkeysvalueof{/data point/z}},
                    \pgfmathprintnumber
                    {\pgfkeysvalueof{/data point/x}},
                    \pgfmathprintnumber
                    {\pgfkeysvalueof{/data point/y}})$%
                }
            }
            ] table[x=c, y=w, z=ped,col sep=comma] {csv/crossing/crossing_1.txt};
        \end{axis}
    \end{tikzpicture}    
\end{subfigure}
\hfill
\begin{subfigure}{0.45\textwidth}
\centering
    \begin{tikzpicture}
        \begin{axis}[
            colormap={seq}{
                color=(red);
                color=(green);
                color=(blue);
            },
            point meta min = 0,
            point meta max = 2,
            point meta={\coordindex},        
            width=\textwidth,
            grid=major,
            view={30}{20},
            zmin=0.0,
            zmax=2.0,
            ztick={0.0,1.0,2.0},
            xtick={0.0,0.5,1.0},
            ytick={0.0,0.5,1.0},
            xmin=0.0,xmax=1.0,
            ymin=0.0,ymax=1.0,
            xlabel={\scriptsize $\probact{c}$},
            ylabel={\scriptsize $\probact{w}$},
            ]         
            \addplot3+[mesh,scatter,only marks,mark size=2pt] table[x=c, y=w, z=ped,col sep=comma] {csv/crossing/crossing_2.txt};
            \addplot3+[no marks,color=white,opacity=1.0,
            nodes near coords={\tiny {\color{black}
                $(\pgfmathprintnumber
                    {\pgfkeysvalueof{/data point/z}},
                    \pgfmathprintnumber
                    {\pgfkeysvalueof{/data point/x}},
                    \pgfmathprintnumber
                    {\pgfkeysvalueof{/data point/y}})$%
                }
            }
            ] table[x=c, y=w, z=ped,col sep=comma] {csv/crossing/crossing_2.txt};
        \end{axis}
    \end{tikzpicture}    
\end{subfigure}

\vspace*{0.2cm}

\begin{subfigure}{0.45\textwidth}
\centering
    \begin{tikzpicture}
        \begin{axis}[
            colormap={seq}{
                color=(red);
                color=(green);
                color=(blue);
            },
            point meta min = 0,
            point meta max = 2,
            point meta={\coordindex},
            title={\small $\mu=4$},
            width=\textwidth,
            grid=major,
            view={30}{20},
            zmin=0.0,
            zmax=2.0,
            ztick={0.0,1.0,2.0},
            xtick={0.0,0.5,1.0},
            ytick={0.0,0.5,1.0},
            xmin=0.0,xmax=1.0,
            ymin=0.0,ymax=1.0,
            xlabel={\scriptsize $\probact{m}$},
            ylabel={\scriptsize $\probact{r}$},
            zlabel={\scriptsize \emph{utility}}
            ]
            \addplot3+[mesh,scatter,only marks,mark size=2pt] table[x=m, y=r, z=car,col sep=comma] {csv/crossing/crossing_4.txt};
            \addplot3+[no marks,color=white,opacity=1.0,
            nodes near coords={\tiny {\color{black}
                $(\pgfmathprintnumber
                    {\pgfkeysvalueof{/data point/z}},
                    \pgfmathprintnumber
                    {\pgfkeysvalueof{/data point/x}},
                    \pgfmathprintnumber
                    {\pgfkeysvalueof{/data point/y}})$%
                }
            }
            ] table[x=m, y=r, z=car,col sep=comma] {csv/crossing/crossing_4.txt};
        \end{axis}
    \end{tikzpicture}    
\end{subfigure}
\hfill
\begin{subfigure}{0.45\textwidth}
\centering
    \begin{tikzpicture}
        \begin{axis}[
            colormap={seq}{
                color=(red);
                color=(green);
                color=(blue);
            },
            point meta min = 0,
            point meta max = 2,
            point meta={\coordindex},
            title={\small $\mu=5$},
            width=\textwidth,
            grid=major,
            view={30}{20},
            zmin=0.0,
            zmax=2.0,
            ztick={0.0,1.0,2.0},
            xtick={0.0,0.5,1.0},
            ytick={0.0,0.5,1.0},
            xmin=0.0,xmax=1.0,
            ymin=0.0,ymax=1.0,
            xlabel={\scriptsize $\probact{m}$},
            ylabel={\scriptsize $\probact{r}$},
            ]
            \addplot3+[mesh,scatter,only marks,mark size=2pt] table[x=m, y=r, z=car,col sep=comma] {csv/crossing/crossing_5.txt};
            \addplot3+[no marks,color=white,opacity=1.0,
            nodes near coords={\tiny {\color{black}
                $(\pgfmathprintnumber
                    {\pgfkeysvalueof{/data point/z}},
                    \pgfmathprintnumber
                    {\pgfkeysvalueof{/data point/x}},
                    \pgfmathprintnumber
                    {\pgfkeysvalueof{/data point/y}})$%
                }
            }
            ] table[x=m, y=r, z=car,col sep=comma] {csv/crossing/crossing_5.txt};
        \end{axis}
    \end{tikzpicture}    
\end{subfigure}


\begin{subfigure}{0.45\textwidth}
\centering
    \begin{tikzpicture}
        \begin{axis}[
            colormap={seq}{
                color=(red);
                color=(green);
                color=(blue);
            },
            point meta min = 0,
            point meta max = 2,
            point meta={\coordindex},  
            width=\textwidth,
            grid=major,
            view={30}{20},
            zmin=0.0,
            zmax=2.0,
            ztick={0.0,1.0,2.0},
            xtick={0.0,0.5,1.0},
            ytick={0.0,0.5,1.0},
            xmin=0.0,xmax=1.0,
            ymin=0.0,ymax=1.0,
            xlabel={\scriptsize $\probact{c}$},
            ylabel={\scriptsize $\probact{w}$},
            zlabel={\scriptsize \emph{utility}}
            ]
            \addplot3+[mesh,scatter,only marks,mark size=2pt] table[x=c, y=w, z=ped,col sep=comma] {csv/crossing/crossing_4.txt};
            \addplot3+[no marks,color=white,opacity=1.0,
            nodes near coords={\tiny {\color{black}
                $(\pgfmathprintnumber
                    {\pgfkeysvalueof{/data point/z}},
                    \pgfmathprintnumber
                    {\pgfkeysvalueof{/data point/x}},
                    \pgfmathprintnumber
                    {\pgfkeysvalueof{/data point/y}})$%
                }
            }
            ] table[x=c, y=w, z=ped,col sep=comma] {csv/crossing/crossing_4.txt};
        \end{axis}
    \end{tikzpicture}    
\end{subfigure}
\hfill
\begin{subfigure}{0.45\textwidth}
\centering
    \begin{tikzpicture}
        \begin{axis}[
            colormap={seq}{
                color=(red);
                color=(green);
                color=(blue);
            },
            point meta min = 0,
            point meta max = 2,
            point meta={\coordindex},        
            width=\textwidth,
            grid=major,
            view={30}{20},
            zmin=0.0,
            zmax=2.0,
            ztick={0.0,1.0,2.0},
            xtick={0.0,0.5,1.0},
            ytick={0.0,0.5,1.0},
            xmin=0.0,xmax=1.0,
            ymin=0.0,ymax=1.0,
            xlabel={\scriptsize $\probact{c}$},
            ylabel={\scriptsize $\probact{w}$},
            ]
            \addplot3+[mesh,scatter,only marks,mark size=2pt] table[x=c, y=w, z=ped,col sep=comma] {csv/crossing/crossing_5.txt};
            \addplot3+[no marks,color=white,opacity=1.0,
            nodes near coords={\tiny {\color{black}
                $(\pgfmathprintnumber
                    {\pgfkeysvalueof{/data point/z}},
                    \pgfmathprintnumber
                    {\pgfkeysvalueof{/data point/x}},
                    \pgfmathprintnumber
                    {\pgfkeysvalueof{/data point/y}})$%
                }
            }
            ] table[x=c, y=w, z=ped,col sep=comma] {csv/crossing/crossing_5.txt};
        \end{axis}
    \end{tikzpicture}    
\end{subfigure}
\caption{Equilibria strategies and utilities for the pedestrian crossing game for different values of $\mu$. The top plot for each value of $\mu$ corresponds to the pedestrian and the plot below to the vehicle. Point coordinates values are $(u,\probact{m},\probact{r})$ for the vehicle and $(u,\probact{c},\probact{w})$ for the pedestrian (where $u$ is the utility).}
\label{fig:eq_cross_alpha_appendix}
\end{figure}
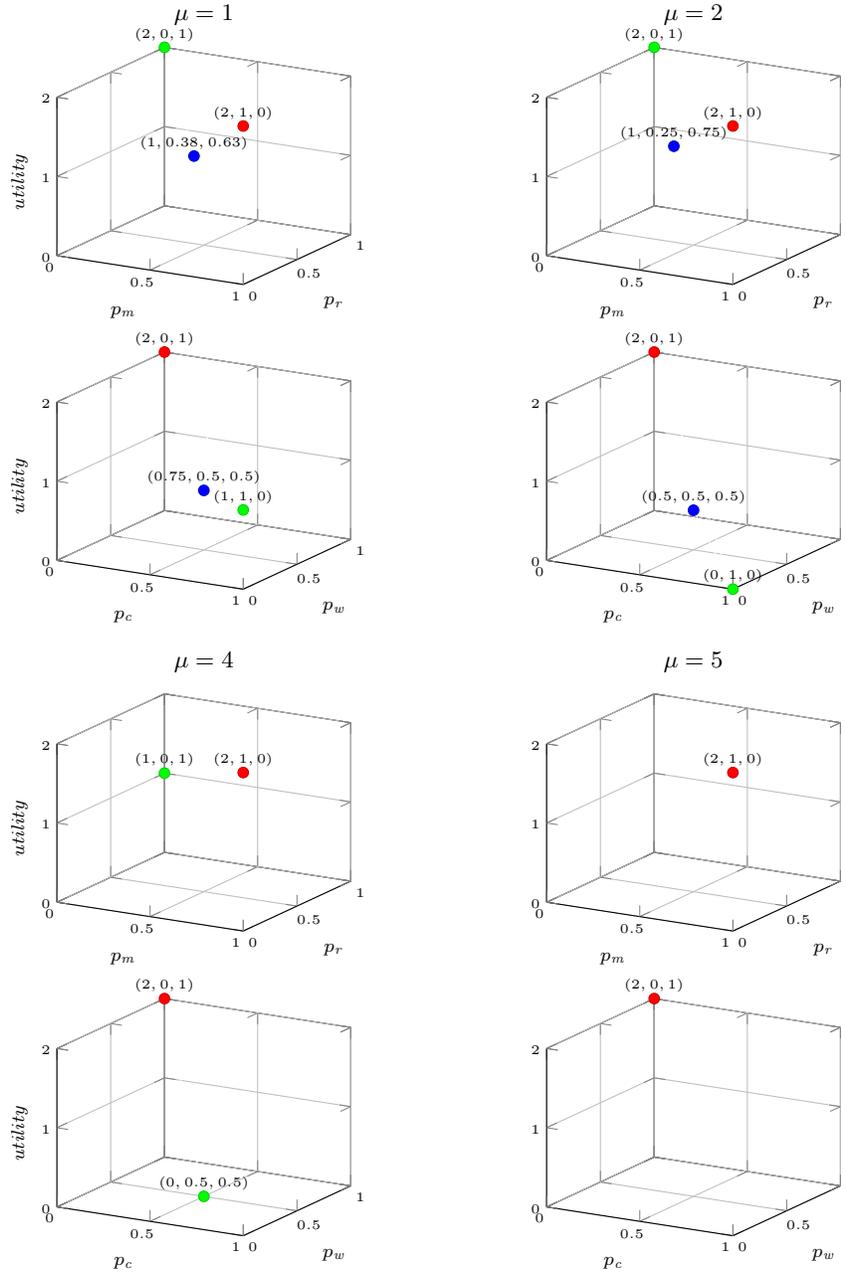

%% file: main.bbl
\begin{thebibliography}{10}
\providecommand{\url}[1]{\texttt{#1}}
\providecommand{\urlprefix}{URL }
\providecommand{\doi}[1]{https://doi.org/#1}

\bibitem{Per24}
Andr\'{e}s, P.: From decision theory to game theory: Reasoning about decisions
  of others (2024), {M}anuscript in preparation, Cambridge University Press.

\bibitem{ANP16}
Aslanyan, Z., Nielson, F., Parker, D.: Quantitative verification and synthesis
  of attack-defence scenarios. In: Proc.\ CSF'16. pp. 105--119. IEEE (2016)

\bibitem{AN08}
Attanasi, G., Nagel, R.: A survey of psychological games: Theoretical findings
  and experimental evidence. In: Innocenti, A., Sbriglia, P. (eds.) Games,
  Rationality and Behavior. Essays on Behavioral Game Theory and Experiments.
  pp. 204--232 (2008)

\bibitem{BCD19}
Battigalli, P., Corrao, R., Dufwenberg, M.: Incorporating belief-dependent
  motivation in games. Journal of Economic Behavior \& Organization
  \textbf{167},  185--218 (2019)

\bibitem{BD09}
Battigalli, P., Dufwenberg, M.: Dynamic psychological games. Journal of
  Economic Theory  \textbf{144}(1),  1--35 (2009)

\bibitem{BD22}
Battigalli, P., Dufwenberg, M.: Belief-dependent motivations and psychological
  game theory. Journal of Economic Literature  \textbf{60}(3),  833--82 (2022)

\bibitem{Tor15}
Bj{\o}rnskau, T.: The zebra crossing game – using game theory to explain a
  discrepancy between road user behaviour and traffic rules. Safety Science
  \textbf{92} (2015)

\bibitem{BRE13}
Brenguier, R.: {PRALINE}: A tool for computing {N}ash equilibria in concurrent
  games. In: Proc.\ {CAV}'13. LNCS, vol.~8044, pp. 890--895. Springer (2013)

\bibitem{DK93}
Dickhaut, J., Kaplan, T.: A Program for Finding Nash Equilibria, pp. 148--166.
  Springer New York, New York, NY (1993)

\bibitem{Elv14}
Elvik, R.: A review of game-theoretic models of road user behaviour. Accident
  Analysis \& Prevention  \textbf{62},  388--396 (2014)

\bibitem{AAD+24}
{Ezzati Amini}, R., Abouelela, M., Dhamaniya, A., Friedrich, B., Antoniou, C.:
  A game-theoretic approach for modelling pedestrian-vehicle conflict
  resolutions in uncontrolled traffic environments. Accident Analysis \&
  Prevention  \textbf{203},  107604 (2024)

\bibitem{FWHT16}
Feng, L., Wiltsche, C., Humphrey, L., Topcu, U.: Synthesis of human-in-the-loop
  control protocols for autonomous systems. IEEE Transactions on Automation
  Science and Engineering  \textbf{13}(2),  450--462 (2016)

\bibitem{GPS89}
Geanakoplos, J., Pearce, D., Stacchetti, E.: Psychological games and sequential
  rationality. Games and Economic Behavior  \textbf{1}(1),  60--79 (1989)

\bibitem{SW03}
Govindan, S., Wilson, R.: A global {N}ewton method to compute {N}ash
  equilibria. Journal of Economic Theory  \textbf{110}(1),  65--86 (2003)

\bibitem{gurobi}
{Gurobi Optimization, LLC}: {Gurobi Optimizer Reference Manual} (2021),
  \href{https://www.gurobi.com}{gurobi.com}

\bibitem{HJK+22}
Hensel, C., Junges, S., Katoen, J.P., Quatmann, T., Volk, M.: The probabilistic
  model checker {Storm}. International Journal on Software Tools for Technology
  Transfer  \textbf{24}(4),  589--610 (2022)

\bibitem{GNP+18}
J.~Gutierrez, J., Najib, M., Perelli, G., Wooldridge, M.: Eve: A tool for
  temporal equilibrium analysis. In: Proc. ATVA'18. LNCS, vol. 11138, pp.
  551--557. Springer (2018), \href{
  https://github.com/eve-mas/eve-parity}{github.com/eve-mas/eve-parity}

\bibitem{JJK+18}
Junges, S., Jansen, N., Katoen, J.P., Topcu, U., Zhang, R., Hayhoe, M.: Model
  checking for safe navigation among humans. In: Proc.\ {QEST}'18. LNCS, vol.
  11024, pp. 207--222. Springer (2018)

\bibitem{KSF+23}
Karabag, M.O., Smith, S., Fridovich-Keil, D., Topcu, U.: Encouraging inferable
  behavior for autonomy: Repeated bimatrix stackelberg games with observations
  (2023)

\bibitem{KSK76}
Kemeny, J., Snell, J., Knapp, A.: Denumerable {M}arkov Chains. Springer (1976)

\bibitem{Kol92}
Kolpin, V.: Equilibrium refinement in psychological games. Games and Economic
  Behavior  \textbf{4}(2),  218--231 (1992)

\bibitem{KNP11}
Kwiatkowska, M., Norman, G., Parker, D.: {PRISM} 4.0: Verification of
  probabilistic real-time systems. In: Proc.\ CAV'11. LNCS, vol.~6806, pp.
  585--591. Springer (2011)

\bibitem{KNPS20}
Kwiatkowska, M., Norman, G., Parker, D., Santos, G.: {PRISM}-games 3.0:
  Stochastic game verification with concurrency, equilibria and time. In:
  Proc.\ {CAV}'20. LNCS, vol. 12225, pp. 475--487. Springer (2020)

\bibitem{KNPS20b}
Kwiatkowska, M., Norman, G., Parker, D., Santos, G.: Multi-player equilibria
  verification for concurrent stochastic games. In: Proc.\ {QEST}'20. LNCS,
  vol. 12289, pp. 74--95. Springer (2020)

\bibitem{KNPS21}
Kwiatkowska, M., Norman, G., Parker, D., Santos, G.: Automatic verification of
  concurrent stochastic systems. Formal Methods in System Design  \textbf{58},
  188--250 (2021)

\bibitem{KNPS22}
Kwiatkowska, M., Norman, G., Parker, D., Santos, G.: Correlated equilibria and
  fairness in concurrent stochastic games. In: Proc.\ TACAS'22. LNCS, vol.
  13244, pp. 60--78. Springer (2022)

\bibitem{LTH87}
Laan, G.V.D., Talman, A.J.J., Heyden, L.V.D.: Simplicial variable dimension
  algorithms for solving the nonlinear complementarity problem on a product of
  unit simplices using a general labelling. Mathematics of Operations Research
  \textbf{12}(3),  377--397 (1987)

\bibitem{LH64}
Lemke, C., J.~Howson, J.: Equilibrium points of bimatrix games. Journal of the
  Society for Industrial and Applied Mathematics  \textbf{12}(2),  413--423
  (1964)

\bibitem{MB18}
Michieli, U., Badia, L.: Game theoretic analysis of road user safety scenarios
  involving autonomous vehicles. In: Proc.\ {PIMRC}'18. pp. 1377--1381 (2018)

\bibitem{PNS04}
Porter, R., Nudelman, E., Shoham, Y.: Simple search methods for finding a
  {N}ash equilibrium. In: Proc.\ AAAI'04. pp. 664--669. AAAI Press (2004)

\bibitem{Rab93}
Rabin, M.: Incorporating fairness into game theory and economics. The American
  economic review pp. 1281--1302 (1993)

\bibitem{SGC05}
Sandholm, T., Gilpin, A., Conitzer, V.: Mixed-integer programming methods for
  finding {N}ash equilibria. In: Proc.\ {AAAI}'05. pp. 495--501. AAAI Press
  (2005)

\bibitem{Sha53}
Shapley, L.: Stochastic games. Proc.\ National Academy of Sciences
  \textbf{39},  1095--1100 (1953)

\bibitem{SKC23}
Shirado, H., Kasahara, S., Christakis, N.A.: Emergence and collapse of
  reciprocity in semiautomatic driving coordination experiments with humans.
  Proc.\ National Academy of Sciences  \textbf{120}(51) (2023)

\bibitem{TGW15}
Toumi, A., Gutierrez, J., Wooldridge, M.: A tool for the automated verification
  of {N}ash equilibria in concurrent games. In: Proc.\ {ICTAC}'15. LNCS,
  vol.~9399, pp. 583--594. Springer (2015)

\end{thebibliography}
